# High-energy side-peak emission of exciton-polariton condensates in high density regime

**Authors:** Tomoyuki Horikiri[1,2,3,4,5,†], Makoto Yamaguchi[4,6,†], Kenji Kamide[3,6], Yasuhiro Matsuo[1,3], Tim Byrnes[7,8,1], Natsuko Ishida[4], Andreas Löffler[9], Sven Höfling[9,10,1], Yutaka Shikano[11,12,13], Tetsuo Ogawa[6,14], Alfred Forchel[9], and Yoshihisa Yamamoto[1,2,3,4,15*]

**Affiliations:**

[1]National Institute of Informatics, Hitotsubashi 2-1-2, Chiyoda-ku, Tokyo 101-8430, Japan.

[2]E. L. Ginzton Laboratory, Stanford University, 348 Via Pueblo, Stanford, California 94305, USA.

[3]The University of Tokyo, 7-3-1 Hongo, Bunkyo, Tokyo 113-8656, Japan.

[4]Center for Emergent Matter Science, RIKEN, 2-1 Hirosawa, Wakoshi, Saitama 351-0198, Japan

[5]Yokohama National University, 79-5 Tokiwadai, Hodogaya-ku, Yokohama, Kanagawa 240-8501, Japan.

[6]Department of Physics, Osaka University, 1-1 Machikaneyama, Toyonaka, Osaka 560-0043, Japan.

[7]New York University, 1555 Century Ave, Pudong, Shanghai, 2002122, China.

[8]NYU-ECNU Institute of Physics at NYU Shanghai, 3663 Zhongshan Road North, Shanghai 200062, China

[9]Technische Physik, Physikalisches Institut and Wilhelm Conrad Röntgen Research Center for Complex Material Systems, Universität Würzburg, Am Hubland, D-97074 Würzburg, Germany.

[10]SUPA, Schoold of Physics and Astronomy, University of St Andrews, KY16 9SS, United Kingdom.

[11]Research Center of Integrative Molecular Systems (CIMoS), Institure for Molecular Science, National Institutes of Natural Sciences, 38 Nishigo-Naka, Okazaki, Aichi 444-8585, Japan.

[12]Institute for Quantum Studies, Chapman University, 1 University Dr., Orange, California 92866, USA.

[13]Materials and Structures Laboratory, Tokyo Institute of Technology, 4259 Nagatsuta, Midori, Yokohama 226-8503, Japan.

[14]Photon Pioneers Center, Osaka University, 2-1 Yamada-oka, Suita, Osaka 565-0871, Japan

[15]ImPACT Program, Japan Science and Technology Agency, 7 Gobancho, Chiyoda-ku, Tokyo 102-0076, Japan.

[*]Correspondence to: yyamamoto@stanford.edu, horikiri@ynu.ac.jp

[†] These authors contributed equally to this work.

## Abstract

In a standard semiconductor laser, electrons and holes recombine via stimulated emission to emit coherent light, in a process that is far from thermal equilibrium. Exciton-polariton condensates -- sharing the same basic device structure as a semiconductor laser, consisting of quantum wells coupled to a microcavity -- have been investigated primarily at densities far below the Mott density for signatures of Bose-Einstein condensation. At high densities approaching the Mott density, exciton-polariton condensates are generally thought to revert to a standard semiconductor laser, with the loss of strong coupling. Here, we report the observation of a photoluminescence sideband at high densities that cannot be accounted for by conventional semiconductor lasing. This also differs from an upper-polariton peak by the observation of the excitation power dependence in the peak-energy separation. Our interpretation as a persistent coherent electron-hole-photon coupling captures several features of this sideband, although a complete understanding of the experimental data is lacking. A full understanding of the observations should lead to a development in non-equilibrium many-body physics.

# Introduction

Semiconductor lasers are one of the most fascinating systems for studying quantum many-body physics in a non-equilibrium regime arising from the interplay of an interacting electron–hole–photon (e–h–p) system. In the context of an e-h-p system in a semiconductor, localized excitons are typically treated as two-level systems with no internal structure and coupled to a continuum of radiation modes confined in a two-dimensional microcavity[1]. On the other hand, at equilibrium without photons, there are several predicted phenomena[2–6] when the electron–hole (e–h) internal structure is taken into account. One example is the e–h Bardeen–Cooper–Schrieffer (BCS) phase in the high e–h density regime[2,7], where the condensation of e–h Cooper pairs opens a gap around the Fermi energy in the electron and hole energy dispersions. Since such BCS physics is based on assumptions of equilibrium in e–h systems, with the complete lack of photons, it has been traditionally conceptually disconnected from semiconductor lasers (e-h-p system).

On the other hand, the phenomenon of exciton-polariton condensation has recently gained tremendous interest[8-11,42], while sharing the same basic structural elements with a semiconductor laser consisting of a number of quantum wells (QWs) coupled to a microcavity structure. The first characteristic difference between semiconductor lasers and exciton-polariton condensates is the lack, or presence respectively, of strong coupling between the bound e-h pairs (excitons) and cavity photons. The strong coupling between the excitons and photons result in a new quasiparticle called the exciton-polariton, which condense into the zero momentum state via stimulated cooling, rather than stimulated emission as in a standard laser[10]. The difference in the mechanism of coherence formation to conventional lasing has suggested that the exciton-polariton system be modelled as a non-equilibrium Bose-Einstein condensate (BEC)[12], with a large number of experiments supporting this interpretation with expected properties such as superfluidity[13]. However, such experiments typically take place at low densities, where the exciton density is several orders of magnitude below the Mott density. At higher densities, conventional wisdom has been that strong coupling is lost, and the system reverts to a standard photon laser[14,15]. The mechanism of this loss of strong coupling is still not very well understood, with only a few theoretical[16,17] and fewer experimental works[18] analysing this regime in detail.

In this study, we directly probe the high density regime of exciton-polariton condensates towards the Mott density. We observe a photoluminescence (PL) sideband in the higher-energy side from the main peak. By the excitation-power dependence of the peak-energy separation, this is different from the upper-polariton peak and cannot be explained by the single-emitter model. The measured PL spectra shows that this is not taken as the conventional semiconductor lasing. Furthermore, we study the PL spectra based on the non-equilibrium e-h-p model including BCS physics, which allows for a treatment in whole regime from non-equilibrium lasing to equilibrium BECs, BCS states and in-betweens. While our measured PL spectra consist of the main peak and the high-energy one, this theory predicts the asymmetric triplet peaks. Our observation has a potential to demonstrate a strong coupling of an electron and a hole under a lasing phase and further leads to deepen non-equilibrium and dissipative many-body physics.

## Results
**Interacting electron-hole-photon (e-h-p) model**

To discuss a high density regime of an exciton-polariton system in a semiconductor microcavity, the BCS physics[19-22] and lasing phase, i.e., an e–h–p system in a non-equilibrium state[22] should be simultaneously discussed. In this section, we qualitatively and quantitatively explain the interplay between the BCS physics and lasing phase from the viewpoint of PL spectra.

In the standard BCS theory without photons, it is well-known that the spontaneous formation of coherence opens an energy gap in the excitation spectrum, which can be described by the BCS gap equation. In the e–h–p system, the corresponding energy gap in the spectrum should exist when the coherence is developed. Energy dispersion of the photon-dressed carriers in the e–h–p system are shown in Fig. 1B. The mechanism generating the energy gap can be understood in analogy with a two-level atomic system (Fig. 1A). Dressed atom states are formed when the cavity photon field[23] and the two-level atom transition are in resonance and strongly coupled (say the coupling strength $g^{JC}$ and the resonance energy $\mu$). The resulting dressed state energy is split into two, where the energy difference between the two energies, $2g^{JC}(n)^{1/2}$ for the total excitation number $n$, corresponds to the rate of the coherent energy transfer between the two-level system and the cavity photon field. As $n$ increases, their coherent energy transfer occurs more rapidly; therefore, the energy splitting becomes large. In the limit of large $n$, the four possible radiative transitions from $n+1$ to $n$ sectors generate three emission peaks at $\mu$ and $\mu \pm \Omega_R$ (two of them degenerate at $\mu$, and $\Omega_R \approx 2g^{JC}(n)^{1/2}$). This is the cavity system version of the Mollow triplet in resonance fluorescence[24,25]. We point out that the Mollow triplet is a direct signature of the coherent coupling between the matter and light fields, i.e. strong coupling. In the present e–h–p system, the corresponding energy gap is observed around the momentum where the valence band with $n+1$ photons and the conduction band with $n$ photons coincide (resonance condition), i.e. when there appears coherence (Fig. 1B). However, the gap formation is not only by the standard Rabi splitting due to the strong light field[23](large $n$), but further assisted by the BCS e–h Coulomb correlations[20–21, 26–28]. In the same way as the standard Mollow triplet with a single atom, the energy gap will show the Mollow-type triplet, which becomes a signature of coherence of the photons with the e-h system and direct evidence of strong coupling.

In what follows, we have employed the recently developed formalism described in Ref. 22 in order to investigate the equilibrium and non-equilibrium nature of the polariton condensate in the high-density regime. In this formalism, electrons (the energy dispersion $\hbar\omega_{e,k}$), holes (the energy dispersion $\hbar\omega_{h,k}$), and photons (the energy dispersion $\hbar\omega_{ph,k}$) are explicitly treated with the Coulomb interactions between carriers ($U_q$) and the light-matter coupling ($g$) within the dipole approximation. One of the advantages of this approach is the applicability to high-density regimes because electrons and holes are explicitly treated with their Coulomb interactions, in contrast to the other approaches assuming excitons as bosonic particles such as dissipative Gross-Pitaevskii equations. By taking into account the pumping baths and the vacuum photon bath, a closed set of equations for the cavity photon field $a_0$, the polarization function $p_k$, and the number of electrons $n_{e,k}$ and holes $n_{h,k}$ can be derived[16,17] within the Hartree-Fock (HF) approximation for the steady state as

$$\partial_t a_0 = 0 = -i\xi_{ph,0} a_0 + ig\sum_k p_k - \kappa a_0, \qquad (1)$$

$$\partial_t p_k = 0 = -i[\tilde{\xi}_{e,k} + \tilde{\xi}_{h,k}]p_k - i\Delta_k N_k - 2\gamma[p_k - p_k^0], \qquad (2)$$

$$\partial_t n_{e(h),k} = 0 = -2\,\text{Im}[\Delta_k p_k^*] - 2\gamma[n_{e(h),k} - n^0_{e(h),k}]. \tag{3}$$

Here, $\hbar\xi_{\text{ph},k} \equiv \hbar\omega_{\text{ph},k} - \mu$ and $\hbar\tilde{\xi}_{e(h),k} \equiv \hbar\omega_{e(h),k} - \sum_{q\neq k} U_{q-k} n_{e(h),k} - \mu/2$ are the (Coulomb-renormalized) energies of the particles measured on a rotating frame with an oscillation frequency $\mu$, $N_k \equiv n_{e,k} + n_{h,k} - 1$ is the population inversion, $\Delta_k \equiv g^* a_0 + \hbar^{-1}\sum_{q\neq k} U_{q-k} p_q$ is the generalized Rabi frequency which represents coherence of the system, $\kappa$ is the photon loss rate, and $\gamma$ is the thermalization rate of the e-h system. Here, we note that $\mu$ can be viewed as an additional unknown variable because the oscillations of $a_0$ and $p_k$ can be eliminated if $\mu$ is appropriately determined. Eq. (1)–(3) are formally the same as the Maxwell–semiconductor–Bloch equations (MSBE) under the relaxation time approximation (RTA) but the major difference is that $p_k^0$ and $n^0_{e(h),k}$ are given by

$$p_k^0 = i\int \frac{d[\hbar\nu]}{2\pi}\left\{G^R_{12,k}(\nu)[1 - f^B_h(-\nu)] - G^R_{21,k}(\nu) f^B_e(\nu)\right\}, \tag{4}$$

$$n^0_{e(h),k} = i\int \frac{d[\hbar\nu]}{2\pi} f^B_{e(h)}(\nu) A_{11(22)}(\pm\nu;k), \tag{5}$$

through the retarded Green's function,

$$G^R_k(\nu) = \begin{pmatrix} \hbar\nu - \hbar\tilde{\xi}_{e,k} + i\hbar\gamma & \hbar\Delta_k \\ \hbar\Delta_k^* & \hbar\nu + \hbar\tilde{\xi}_{h,k} + i\hbar\gamma \end{pmatrix}^{-1}, \tag{6}$$

the single particle spectral function,

$$A_{\alpha\alpha'}(\nu;k) = i[G^R_{\alpha\alpha',k}(\nu) - G^{R*}_{\alpha'\alpha,k}(\nu)], \tag{7}$$

and the Fermi distribution function in the electron (hole) pumping bath, $f^B_{e(h)}(\nu) \equiv [\exp\{\beta[\hbar\nu - (\mu^B_{e(h)} - \mu/2)]\} + 1]^{-1}$, with the chemical potential $\mu^B_{e(h)}$ and inverse temperature $\beta$ ($\equiv k_B T$). As a result, if $\omega_{e,k} = \omega_{h,k}$ is assumed with $\mu^B_e = \mu^B_h$, and $\gamma$ and $T$ are assumed to be small for simplicity (all assumed hereafter unless otherwise specified), the above equations result in the BCS gap equation when (I) $\min[2\hbar E_k] \gtrsim \mu_B - \mu$ (quasi-equilibrium) with $\mu_B \equiv \mu^B_e + \mu^B_h$. In contrast, the MSBE under the RTA, which describes semiconductor lasers, can be recovered when (II) $\mu_B - \mu \gtrsim \min[2\hbar E_k]$ (lasing; non-equilibrium). This formalism is referred to as the BEC-BCS-LASER crossover theory within the HF approximation.

Here we find from Eq. (6) and (7) the single particle spectral functions for electrons (holes) are given by

$$A_{11(22)}(\nu;\mathbf{k}) = 2|u_\mathbf{k}|^2 \frac{\hbar\gamma}{[\hbar\nu \mp \hbar E_\mathbf{k}]^2 + [\hbar\gamma]^2} + 2|v_\mathbf{k}|^2 \frac{\hbar\gamma}{[\hbar\nu \pm \hbar E_\mathbf{k}] + [\hbar\gamma]^2}, \tag{8}$$

having peaks at $\nu = \pm E_k$ given by

$$E_k \equiv \sqrt{[\tilde{\xi}^+_{\text{eh},k}]^2 + |\Delta_k|^2}, \tag{9}$$

and weights, $|u_k|^2$ and $|v_k|^2$ respectively, given by the Bogoliubov coefficients,

$$u_k \equiv \sqrt{\frac{1}{2} + \frac{\tilde{\xi}^+_{\text{eh},k}}{2E_k}}, \quad v_k \equiv e^{i\theta_k}\sqrt{\frac{1}{2} - \frac{\tilde{\xi}^+_{\text{eh},k}}{2E_k}}, \tag{10}$$

where $\tilde{\xi}^+_{\text{eh},k} \equiv [\tilde{\xi}_{e,k} + \tilde{\xi}_{h,k}]/2$ and $\theta_k \equiv \arg(\Delta_k)$. In Eq. (8)–(10), it is important to notice the remarkable similarities to the standard BCS theory describing superconductors. These single particle spectral functions describe the renormalized single particle energies, and it is well-known that gaps are opened around the energy $\pm\mu/2$ with the magnitude of min$[2\hbar E_k]$ (the minimum of $2\hbar E_k$ when the wavenumber $k$ is scanned). The picture shown in Fig. 1 (B) can thus be obtained, and one can now notice that the Mollow's dressed picture in semiconductor is conceptually connected to the gaps in the BCS theory. However, we have to emphasize that the unknown variables in Eq. (8)–(10) are determined by the BEC-BCS-LASER crossover theory rather than the BCS theory. The PL spectra shown later can be roughly discussed with the knowledge of the steady state populations in Eqs.(1)-(3), and the single-particle spectral functions (with the energy dispersion $E=\pm E_k$) in Eq. (8), while they are calculated from the photon Green's function in the practical numerical simulation (see Supplementary Information S3.3).

**Experiment**
The time evolution of the PL spectra of the high density exciton-polariton condensates after a pulse excitation is studied by using a streak camera as shown in Fig. 2. Figure 2A is an example of the time-resolved PL spectra obtained at strong pump power ($P/P_{th}$=75) as a function of time after a triggering pulse's arrival at the streak camera, where the strong PL signal is observed at around an instant (Time~100 ps) with a strong main peak at 1.612 eV and an extra high-energy peak at 1.622 eV (the PL spectra at the instant is shown in Fig.2B). The strong emission is followed by a relaxation decay of a hundred picosecond timescale. During the decay processes, the emission intensity decreases, while the main emission peak energy gradually decreases and is considered to approach finally to that of the lower polariton (LP) ground state (The decay process is focused in Fig. S9 in the Supplementary Information).

A remarkable finding, which is the main focus of this paper, is the emergence of an extra high-energy side peak, which is found only under strong pumping far above the condensation threshold, $P/P_{th} \gtrsim 20$. This is clearly seen in Fig.2C, a collection of such time-resolved PL spectra (at the instant of strong emission, similarly to Fig. 2B) taken at various pump power, 0.6< $P/P_{th}$ <340. Blue shift is observed in the energies of the two peak emissions (main and side) as the pump power increases, and the main peak approaches the bare cavity photon energy (or slightly above it) at the highest pump power. The feature found in the main emission peak is consistent with past predictions of a high density polariton state[1,19–21, 30] and also with our simulation. As for this high-energy side peak, the emission energy increases with the pump power, and is clearly different from that of the upper polariton (UP) since it gradually evolves from the main peak energy (as predicted for cavity system in Ref. 24). The pump power dependence of the energy separation from the main peak is shown in Fig. 2D. In contrast to the main peak, these observations as for the high-energy side peak require further explanation, since it is far beyond the prediction of literature; it is widely believed that the polariton condensate change its nature to conventional photon lasing at high density, and conventional photon lasing does not result in such a high-energy side peak. The deviation from the conventional photon lasing has also been supported by the temperature dependence of the PL in the same sample[18].

This side peak with the pump-power dependent peak separation reminds us of Mollow triplet spectra in coherently driven two-level emitters discussed above (resonance fluorescence). Actually, the Mollow-triplet side-peak separation is known to have a square-root dependency on the pump power which is not too far from our observation (the square root dependency is shown by a black solid line in Fig. 2D). Of course, a considerable deviation from the square root dependency is not surprising as it is also seen in the theoretical results in the Supplementary Material. Before all, our emission sources are semiconductor carriers much more complicated than single two-level atom systems due to e.g. the dispersion, dephasing, higher-order Coulomb effects including carrier-induced relaxation, and carrier heating etc. However, the most mysterious deviation from the conventional resonance fluorescence is that the low-energy side peak theoretically predicted[21,22,24] is missing at any excitation power in our experiments.

**Comparison between experiment and theory**

Now, let's see what predictions on the PL spectra can be drawn from the simulation by our interacting e-h-p model, and then, compare them with our observation of the high-energy side peak emission.

Our theory predicts that the system is in quasi-equilibrium in the low-excitation regime when the quasi-equilibrium condition (I) is satisfied. This condition, (I) the gap energy (=min[$2\hbar E_k$]) is larger than the difference $\mu_B - \mu$, means that the Fermi level of the pumping bath does not exceed the energy gap as shown in Fig. 3A. In this case, it is expected that the high-energy peak in the triplet does not become bright since the carriers cannot be supplied to the renormalized-band states above the gap (namely, the emission channel at $E>\mu$+min[$2\hbar E_k$] is closed). In contrast, in the high-excitation regime, the e–h–p system behaves differently from it would in the low-excitation regime, and the system enters the non-equilibrium regime when another condition (II) is satisfied; the difference $\mu_B - \mu$ becomes larger than the gap energy, min[$2\hbar E_k$]. In this case, the pumping baths can supply the carriers above the energy gap (Fig. 3B). Once this condition is satisfied, the emission channel at $E>\mu$+min[$2\hbar E_k$] would be opened and the high-energy side peak emission begins to occur. These expectations are indeed confirmed by the simulated PL spectra shown in Fig. 4 (A) and Fig. 4 (B), respectively, under the quasi-equilibrium (I) and non-equilibrium (II) conditions (the small but nonzero high-energy side peak intensity in Fig. 4 (A) is due to the non-zero thermal/quantum fluctuations). From this observation, a short conclusion from our e-h-p model is drawn; the bright high-energy side peak emission occurs only in the non-equilibrium condition.

We note, however, that many-body effects also play an important role in the simulated PL spectra. In semiconductor materials (with no cavity), it is well known that e–h Coulomb interactions can cause significant enhancement of the PL intensity around the Fermi edge at low temperature, even for a plasma state[31, 32]. As found in our previous studies on the carrier population and optical gain spectra[16,17], this effect was shown to survive also in our case (Fig. 4 (B)). Furthermore, the Fermi-edge enhancement becomes more pronounced when the difference of the Fermi energies (~$\mu_B$) roughly coincides with the energy separation ($\mu$ + min[$2\hbar E_k$]) between the upper and lower edges of the gaps (see also Fig. 3)[17]. As a result, the high-energy peak is further enhanced and exceeds the low energy peak when $\mu_B - \mu \sim$ min[$2\hbar E_k$] in our calculations. It is also instructive to note that $\mu_B - \mu \sim$ min[$2\hbar E_k$] is satisfied at the pump power where the system crosses over from the quasi-equilibrium states into lasing states[16]. Therefore, we can draw the second conclusion; the PL spectra exhibits asymmetric triplets with stronger high-energy side peak (than the low-energy side peak) near the crossover regime into lasing (see also Fig. 9(d) of Ref.17). We note that such asymmetric PL spectra are never obtained for non-interacting two-level atom models (see also the Supplementary Information and Fig. S6), showing an impact of the many-body effect present in highly-excited e-h-p systems. By seeing the result, we have to recognize that our theoretical results show a large quantitative discrepancy from the measured PL spectra, since the latter exhibit only the main and high-energy side peak emissions with the missing low-energy side-peak emission.

However, it is also interesting to notice a qualitative similarity between the theory and experiment, the non-monotonic pump-power dependence of the high-energy peak separation (theory: blue points in Fig. 4C, and experiments: Fig.2D); the energy separation (i.e. the gap, min[$2\hbar E_k$], in theory) increases with pump power, but decreases eventually at too strong pumping. If the gap is regarded as the binding energy of e-h pairs, the simulation result also implies that the e-h pair binding survives at strong pumping above the crossover into lasing[16], and gradually quenches with increasing pump power where the e-h plasma is eventually formed.

Next, let us study whether the non-equilibrium condition, which is the requirement for the high-energy side peak to be found in our theory, is satisfied or not in the experiments, especially in the high-excitation regime. For this purpose, in Figure 5A, we show the pump-power dependence of the PL intensity obtained in experiments. The second nonlinear increase in the PL intensity as reported in various studies[9,14,15,33-35] is not seen in our experiments, even at our highest available pump power. This also implies that conventional photon lasing with e–h plasma gain does not occur here[18]. A discrepancy exists between the experimental and theoretical PL intensities (Fig. 5A and Fig. 5B) regarding whether the second nonlinear increase occurs. We should discuss this point more carefully from the theoretical viewpoint; the simulation results depend on the parameters; roughly speaking, the second nonlinear increase (namely, the second threshold) is less visible for larger broadening factors by increasing $T$, $\gamma$, and $\kappa$, which is understood by the precise criteria (including these parameters) for the quasi-equilibrium/non-equilibrium conditions; there is a crossover regime between them, i.e. min[$2\hbar E_k$] − ($2\hbar\gamma$ +$2k_BT$) ≲ $\mu_B$ − $\mu$ ≲ min[$2\hbar E_k$] + ($2\hbar\gamma$ +$2k_BT$), and therefore, the threshold feature becomes clear only if the crossover regime is negligibly small. We verified this understanding by simulating the temperature dependence of the PL intensity (not shown). The third conclusion drawn from this consideration is that the second nonlinear increase of the PL intensity cannot be a good clue indicating that the system transitions into the non-equilibrium regime. Here, we would like to add as well that the second nonlinear increase is not necessarily related to the e–h pair breaking, as already shown in Ref. 16.

In order to clarify whether the e-h-p systems were in the quasi-equilibrium or non-equilibrium conditions in our experiments, the dependence of the PL intensity on the PL energy of the main peak ($\mu$ in theory) is shown in Fig. 5C and 5D. As shown above, under the quasi-equilibrium condition ($\mu_B$ − $\mu$ ≲ min[$2\hbar E_k$]), the e-h-p system is described within the thermal equilibrium theory, where the main peak energy $\mu$ is regarded as the chemical potential of the e-h-p systems. In this case, the cavity photon number should diverge when $\mu$ approaches the bare cavity energy[20] (the right vertical dashed line) as plotted by a black line in Fig. 5D. This divergence originates from the basic physics of thermal-equilibrium theory of bosons, which are photons in our case. This prediction leads to a criterion to distinguish between the quasi-equilibrium and non-equilibrium states; if one finds non-divergent PL intensity with the main emission energy (=$\mu$) reaching the bare cavity energy, it gives a proof to show the system is in the non-equilibrium. In the theoretical result (filled circles in Fig. 5D), a large discrepancy from the thermal equilibrium result (a black line) is found as $\mu$ approaches the bare cavity energy, where the system is found to be in the non-equilibrium. The beginning point of this discrepancy ($\mu$ − $E_{cav}$ ~ −4 meV) indicates the crossover regime between the quasi-equilibrium and non-equilibrium, which fully agrees with that obtained directly from the condition $\mu_B$ − $\mu$ ~ min[$2\hbar E_k$] (the two different regimes indicated by the red and blue plots in Fig. 5D). The experimental data in Fig. 5C clearly shows non-divergent PL intensity even when the main peak energy ($\mu$) reaches the bare cavity energy. Therefore, we can safely state that the observed PL data taken at the regime ($\mu \sim E_{cav}$) come from the non-equilibrium states of the e-h-p system. Furthermore, we show the region containing a high-energy side peak in the PL spectra by the blue shaded areas for both the experiments (Fig. 5C) and theory (Fig. 5D). Especially for the experimental results in Fig. 5C, the shaded area with the high-energy side peak emission largely overlaps with the theoretically supported non-equilibrium regime ($\mu \geq E_{cav}$). This consistency partly supports our theoretical interpretation of the high-energy side peak. However, by comparing Fig. 5C with Fig. 5D, we have to stress that the theoretical and experimental data do not fit quantitatively.

## Discussion

In conclusion, we have performed a study of high density exciton-polariton condensates towards the Mott density and observed a high-energy sideband PL. We have compared this to a theory of non-equilibrium e-h-p system, generalizing polariton BCS theory to the non-equilibrium regime. After the comparison, several disagreements between our experiment and theory still exist. We point out that this theory is only one possible explanation of the physics observed in our experiment. Further work would be required to show that the observations are consistent with other possible explanations, some of which are discussed below.

Here, we add some discussions on the discrepancy between the theory and experiments, and mention some important factors and the effects which are neglected in our model. One is the observed relaxation dynamics after a pulse excitation (Fig.2A), while the theory is dealing with the stationary state. Therefore, whether the stationary condition is fulfilled remains questionable. If the relaxation time to the (transient) stationary state (corresponding to Time~100 ps in Fig.2A) is taken into account, the high-energy peak emission could be enhanced, since the carriers initially supplied from the high-energy region. Dephasing omitted in our mean-field theory is known to enhance the emission from off-resonant modes (the side-peak emission in our case) in cavity-QED research[36-39]. It is also well-known to destroy coherence and reduce the gap (the peak separation in our case) in the condensed matter research. Besides, the spontaneous emission from the quantum wells directly into free space was also neglected in our theory. Taking all these effects into account in the theory might reduce the large quantitative discrepancy from the experimental results, which is far beyond the scope of this paper.

One may seek for the origin of the high-energy side peak by other explanations different from our coherent e-h-p coupling scenario. In particular, the single-emitter Mollow triplet in the presence of detuning and dephasing has been shown to give an asymmetric Mollow spectrum[40]. However, our experiments were performed at high densities towards the Mott density. Therefore, it is essential to take into account of the underlying Fermionic nature of the electrons and holes together with their Coulomb interaction. While a single-emitter Mollow triplet under suitable conditions may have superficial similarities, we believe that it is less plausible than the theoretical analysis we have presented in this work.

Another scenario would be for instance that the upper energy peak could be the band to band transition, with the lower energy peak being the bare cavity mode. We believe that it is unlikely that the band-edge emission account for the upper energy peak as the renormalized band edge is lower in energy than the cavity resonance (= QW exciton resonance) in this high density regime[26]. In contrast, the Fermi-edge emission, which was actually observed in Kim *et al.*[41] in highly-excited semiconductor QWs without coupling to a cavity by using a streak camera, can be a possible candidate for our upper energy peak. For this scenario, the Fermi-edge should lower as time proceeds due to the radiation decay, hence, the emission energy should show a red shift, e.g., from 1.622 eV to 1.612 eV in Fig. 2A. In this case, there are two possibilities: (i) the red shift of the Fermi-edge emission energy is very fast and occurs within the time resolution of our streak camera. This could be possible because the radiative decay rate should be higher than Kim *et al*. in our system with the microcavity and higher excitation density. (ii) the red shift of the emission energy occurs in a time scale longer than the time-resolution. In the former case (i), the observed PL spectra at the emission time should consist of a strong emission at cavity energy 1.612 eV plus a broad tail spread between 1.622 eV and 1.612 eV, which is quite different from the observed PL spectra with clear two peaks (e.g. Fig. 2B). Thus, this possibility can be safely excluded. In the latter case (ii), the red shift should have been observed by our streak camera. However, our experimental data does not exhibit such red shift (as in Fig. 2A), at any pump power with a well-defined side peak. Therefore, the latter possibility is also excluded. In either case, the Fermi-edge emission does not account for the observed high-energy side peak.

On the other hand, this absence of the red shift of the high-energy peak (as in Fig. 2A) does not contradict with our coherent e-h-p coupling scenario. In our theory, the high-energy peak emission occurs only if the Fermi-edge ($=\mu_B - \mu$, measured from the main emission) is close to the gap energy, whereas the gap energy robustly stays at the same energy position as long as the strong main peak emission exists (the intensity $\sim \kappa |a_0|^2$). As time proceeds, the Fermi-edge is quickly lowered and detuned from the gap energy. This violates the requirement for the emission to occur, resulting in a sudden quench of the upper energy emission as found in Fig. 2A.

**References:**
1. Keeling, J. Eastham, P. R. Szymanska, M. H. and Littlewood, P. B., BCS-BEC crossover in a system of microcavity polaritons. *Phys. Rev. B* 72, 115320 (2005).
2. Keldysh, L. V. Kopaev, Y. V. Possible instability of the semimetallic state toward Coulomb interaction. *Sov. Phys. Solid State* 6, 2219 (1965).
3. Comte, C. Nozières, P. Exciton Bose condensation: the ground state of an electron-hole gas I. Mean field description of a simplified model. *J. Physique* 43, 1069 (1982).
4. Pieri, P. Neilson, D. and Strinati, G. C., Effects of density imbalance on the BCS-BEC crossover in semiconductor electron-hole bilayers. Phys. Rev. B **75**, 113301 (2007).
5. Yamashita, K. Asano, K. and Ohashi, T. Quantum condensation in electron–hole bilayers with density imbalance. *Journal of Physical Society of Japan* **79**, 033001 (2010).
6. Parish, M. M. Marchetti, F. M. and Littlewood, P. B. Supersolidity in electron-hole bilayers with a large density imbalance. *Euro. Phys. Lett.* **95**, 27007 (2011).
7. Versteegh, M. van Lange, A. J. Stoof, H. T. C. and Dijkhuis, J. I., Observation of preformed electron-hole Cooper pairs in highly excited ZnO, *Phys. Rev. B* **85**, 195206 (2012).
8. Kasprzak, J. *et al*., Bose–Einstein condensation of exciton polaritons. *Nature* **443**, 409 (2006)
9. Deng, H. *et al*., Condensation of semiconductor microcavity exciton polaritons. *Science* **298**, 199 (2002).
10. Deng, H., Haug, H. & Yamamoto, Y. Exciton-polariton Bose-Einstein condensation. Rev. Mod. Phys. 82, 1489 (2010).
11. Carusotto, I. & Ciuti, C. Quantum fluids of light. Rev. Mod. Phys. 85, 299 (2013).

12. Wouters, M. & Carusotto, I. Excitations in a nonequilibrium Bose-Einstein condensate of exciton polaritons. Phys. Rev. Lett. 99, 140402 (2007).
13. Amo, A. et al. Superfluidity of polaritons in semiconductor microcavities. Nat. Phys. 5, 805 (2009).
14. Dang, L. S. Heger, D. Boeuf, F. and Romenstain, R., Stimulation of polariton photoluminescence in semiconductor microcavity. *Phys. Rev. Lett.* **81**, 3920 (1998).
15. Tempel, J. S. *et al.*, Characterization of two-threshold behavior of the emission from a GaAs microcavity. *Phys. Rev. B* **85**, 075318 (2012).
16. Yamaguchi, M. *et al.*, Second thresholds in BEC-BCS-Laser crossover of exciton-polariton systems. *Phys. Rev. Lett.* **111**, 026404 (2013).
17. Yamaguchi, M *et al.*, Generating functional approach for spontaneous coherence in semiconductor electron-hole-photon systems. Phys. Rev. B **91**, 115129 (2015).
18. Horikiri, T. *et al.*, Temperature Dependence of Highly Excited Exciton Polaritons in Semiconductor Microcavities. *J. Phys. Soc. Jpn.* **82**, 084709 (2013).

19. Szymanska, M. H. Keeling, J. and Littlewood, P. B., Non-equilibrium quantum condensation in an incoherently pumped dissipative system *Phys. Rev. Lett.* **96**, 230602 (2006).
20. Kamide, K. and Ogawa, T. What determines the wave function of electron-hole pairs in polariton condensates? *Phys. Rev. Lett.* **105**, 056401 (2010).
21. Byrnes, T. Horikiri, T. Ishida, N. and Yamamoto, Y., A BCS wavefunction approach to the BEC-BCS crossover of exciton-polariton condensates. *Phys. Rev. Lett.* **105**, 186402 (2010).
22. Yamaguchi, M. Kamide, K. Ogawa, T. and Yamamoto, Y., BEC–BCS-laser crossover in Coulomb-correlated electron–hole–photon systems. *New. J. Phys.* **14**, 065001 (2012).
23. Mollow, B. R., Power spectrum of light scattered by two-level systems. *Phys. Rev.* **188**, 1969 (1969).
24. Sanchez-Mondragon, J. J. Narozhny, N. B. and Eberly, J. H., Theory of spontaneous-emission line shape in an ideal cavity, *Phys. Rev. Lett.* **51**, 550 (1983).
25. Quochi, F. *et al.*, Strongly driven semiconductor microcavities: From the polariton doublet to an ac Stark triplet. *Phys. Rev. Lett.* **80**, 4733 (1998).
26. Haug, H. and Koch, S. W., *Quantum Theory of the Optical and Electronic Properties of Semiconductors* 5th ed. (Singapore: World Scientific, 2009).
27. Schmitt-Rink, S. Chemla, D. S. and Haug, H., Non-equilibrium theory of the optical Stark effect and spectral hole burning in semiconductors. *Phys. Rev. B* **37**, 941 (1988).
28. Henneberger, K. and Haug, H., Nonlinear optics and transport in laser-excited semiconductors. *Phys. Rev. B* **38**, 9759 (1988).
29. Roumpos, G. *et al.*, Single vortex–antivortex pair in an exciton-polariton condensate. *Nature Phys.* **7**, 129 (2011).
30. Ishida, N. *et al.*, Photoluminescence of high-density exciton-polariton condensates. *Phys. Rev. B* **90**, 241304(R) (2014).
31. Schmitt-Rink, S. Ell, C. and Haug, H., Many-body effects in the absorption, gain, and luminescence spectra of semiconductor quantum-well structures. *Phys. Rev. B* **33**, 1183 (1986).
32. Skolnick, M. S. *et al.*, Observation of a many-body edge singularity in quantum-well luminescence spectra. *Phys. Rev. Lett.* **58**, 2130 (1987).

33. Bajoni, D. *et al.*, Polariton laser using single micropillar GaAs-GaAlAs semiconductor cavities. *Phys. Rev. Lett.* **100**, 047401 (2008).
34. Nelsen, B. Balili, R. Snoke, D.W. Pfeiffer, L. and West, K., Lasing and polariton condensation: Two distinct transitions in GaAs microcavities with stress traps. *J. Appl. Phys.* **105**, 122414 (2009).
35. Tsotsis, P. *et al.*, Lasing threshold doubling at the crossover from strong to weak coupling regime in GaAs microcavity. *New J. Phys.* **14**, 023060 (2012).
36. Hennessy, K. *et al.*, Quantum nature of s strongly coupled single quantum dot-cavity system. *Nature* **445**, 896 (2007).
37. Winger, M. *et al.*, Explanation of Photon Correlations in the Far-Off-Resonance Optical Emission from a Quantum-Dot–Cavity System. *Phys. Rev. Lett.* **103**, 207403 (2009).
38. M. Settnes, P. Kaer, A. Moelbjerg, and J. Mork, Auger Processes Mediating the Nonresonant Optical Emission from a Semiconductor Quantum Dot Embedded Inside an Optical Cavity. *Phys. Rev. Lett.* **111**, 067403 (2013).
39. Yamaguchi, M. Asano, T. and Noda, S., Third emission mechanism in solid-state nanocavity quantum electrodynamics. *Rep. Prog. Phys.* **75**, 096401 (2012).
40. Valle, E. and Laussy, F. P., Regime of strong light-matter coupling under incoherent excitation. *Phys. Rev.* A **84**, 043816 (2011).
41. Kim, J. H. *et al.*, Fermi-edge superfluorescence from a quatum-degenerate electron-hole gas. *Sci. Rep.* **3**, 3283 (2013).
42. Byrnes, T, Kim, N. Y., and Yamamoto, Y., Exciton-polariton condensates. Nature Phys. **10**, 803 (2014).

**Methods**: We used an AlAs/AlGaAs distributed Bragg reflector microcavity sample in which 12 GaAs quantum wells (QWs) are embedded at central antinodes of the cavity photon field[29,18]. The number of top (bottom) layers used to obtain the PL from the top surface is 16 (20). The 12 QWs are divided into 3 groups and positioned at the 3 highest mode intensity antinodes of the microcavity. The detuning between the microcavity photon energy and the QW exciton at around 1.612 eV is close to zero at the 8 K temperature of the present work; the normal mode splitting at zero in-plane momentum is 14 meV. A mode-locked Ti:Sapphire laser with a 76-MHz repetition rate and an energy around 1.67 eV is used as the pump laser to utilize its 3-ps pulse width for the above-band excitation. The laser injection angle into the sample is around 50−60° from the normal which corresponds to $k \sim 7 \times 10^4$/cm[18]. The pump laser energy is set to maximize the injection rate into the sample reflection dip due to the cavity photon mode. The PL from the sample is focused onto the entrance slit of a spectrometer attached in front of a streak camera or a time-integrated CCD camera by an objective lens and subsequent lenses, and then is reflected at the grating to extract PL energy. In this study, the central area (about 1.2 μm × 1μm) of the pump laser spot with ~50 μm diameter is selected by the horizontal and vertical slits of the spectrometer and the streak camera. Hence, the outside of the central area with significantly different pumping intensities do not affect the observed signal.

**Supplementary Information** is available in the online version of the paper.

**Acknowledgments**: The authors thank P. Littlewood, J. Keeling, I. Carusotto, M. Kuwata-Gonokami, and M. Nomura for their helpful comments. This research was supported by JSPS through its FIRST Program and KAKENHI Grant Numbers 20104008, 24740277, 25800181, 26287087 and 26790061, Space and Naval Warfare Systems (SPAWAR) Grant N66001-09-1-2024, the Ministry of Education, Culture, Sports, Science, and Technology (MEXT), the Project for Developing Innovation Systems of MEXT, the National Institute of Information and Communications Technology (NICT), the joint study program at the Institute for Molecular Science, the Shanghai Research Challenge Fund, NSFC, and NTT Basic Laboratories.

**Author contributions**
T. H. and M. Y. equally contributed to this work. T. H. performed optical experiments. M. Y. performed theoretical calculations. A. L., S.H. and A. F. grew and fabricated the sample. T. H. and Y. Y. designed the experiment. M.Y, K. K., T. B., N. I., Y. S. and T. O. prepared the theoretical work. T. H. and Y. M. analysed data. T. H., M. Y., K. K., Y. S., and T. B. prepared the manuscript. T. O., S. H. and Y.Y. led the discussion. Y. Y. organized this work.

**Additional information**
Competing financial interests: The authors declare no competing financial interests.

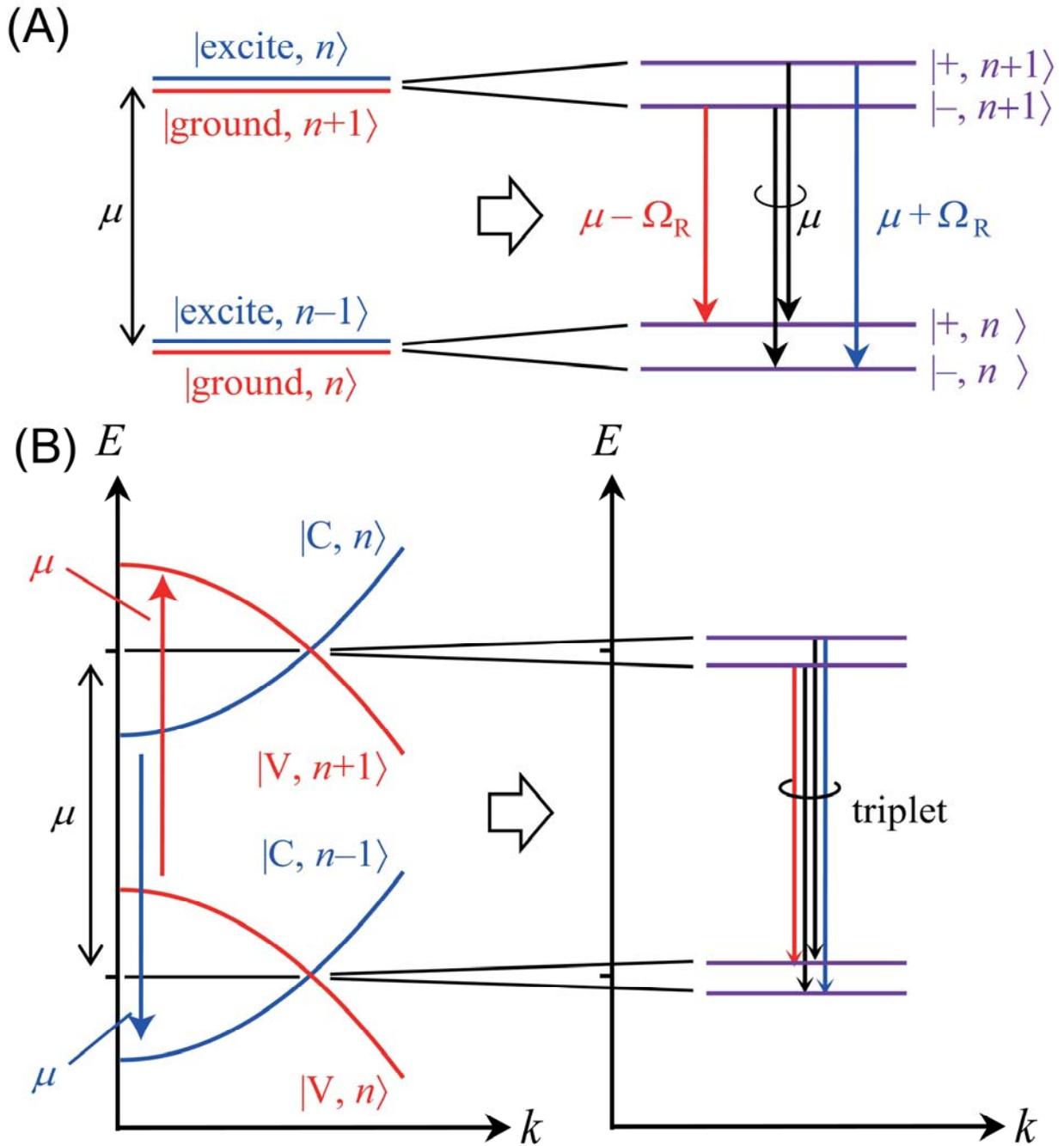

**Fig. 1|** Energy diagrams of the dressed states in the two level emitter (A) and the e–h–p dispersion (B). In the left panel of (B), the dipole coupling to the cavity photons and the e–h attractive Coulomb interactions are neglected, while it is include in the right panel. In this case, the electron band (the solid blue curve) is mixed with the $+\hbar\omega_0$-shifted hole band (the dashed red curve). In the same manner, the hole band (the solid red curve) is mixed with the $-\hbar\omega_0$-shifted electron band (the dashed blue curve). The triplet spectrum is formed in a certain wavenumber regime, where the valence band of the $n+1$ total excitation numbers and the conduction band of the $n$ coincide.

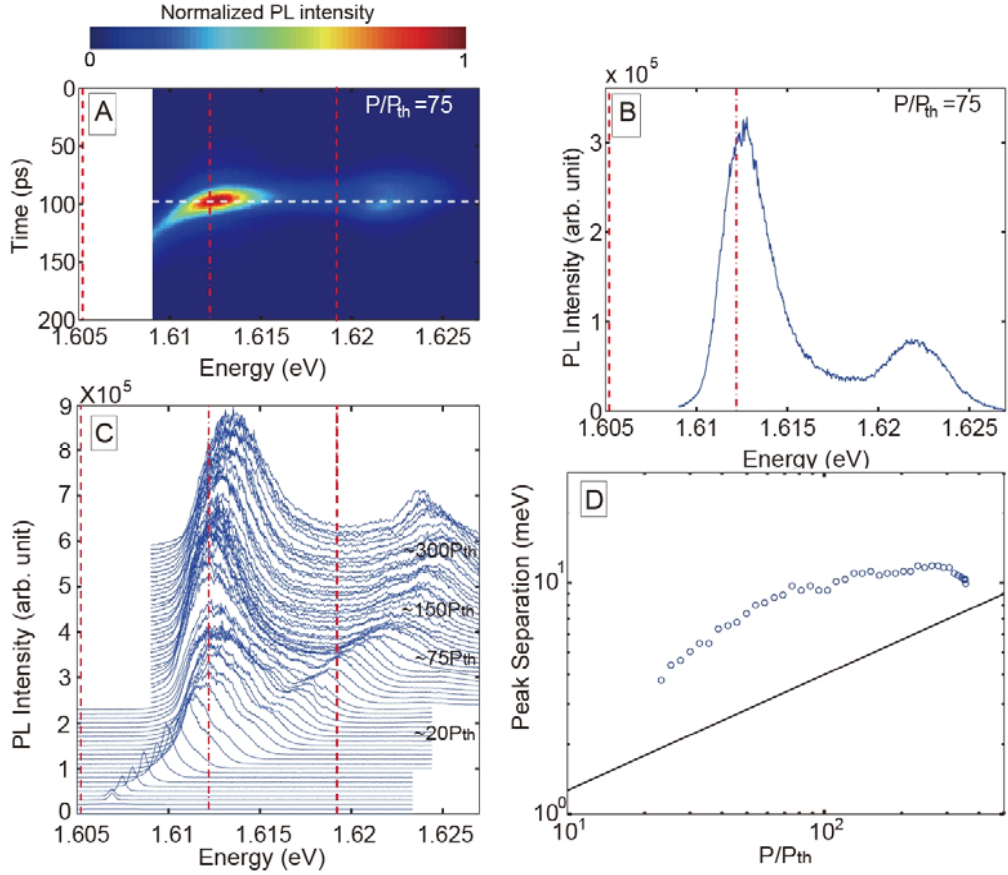

**Fig. 2|** Time-resolved spectroscopy of the photoluminescence. (**A**) PL spectrum of excitation density at $75P_{th}$, where the threshold pump laser power $P_{th}$ of 2.9 mW is determined from the nonlinear PL increase of the LP ground state. Horizontal and vertical axes represent energy and time, respectively. The origin of the time axis is not the pump laser pulse arrival time but the time of the trigger pulse arrival at the streak camera. (**B**) Cross-section of PL spectrum at the horizontal dotted line of (A) giving the maximum PL intensity. Vertical axis is PL intensity while horizontal axis corresponds to PL energy. (**C**) Excitation density dependence of the PL spectra from 0.6 $P_{th}$ to 340 $P_{th}$. The time of the maximum PL intensity at each excitation density is extracted. As pump power increases, increasing offsets are added to the spectra to allow them to be distinguished. The lowest curves corresponding to PL below $P_{th}$ are much weaker, therefore, they are shown as almost horizontal lines The red dot-dashed line and the red dashed lines represent the energy of the cavity photon mode (1.612 eV) and the energies of the ground states of the UP (1.619 eV) and LP (1.605 eV) far below $P_{th}$, respectively. The photosensitive area of the CCD camera attached to our streak camera limits the observed width. Therefore, as high-energy peak shifts to higher-energy as pump power increases, plotted energy area changes. (**D**) Experimental energy separation between the two peaks at the instance of maximum PL (as in Fig. 2B). Horizontal axis shows the pump power normalized by the power giving the condensation threshold. The two peaks are visible above around 20 $P_{th}$ as seen in Fig. 2C. Therefore, Fig. 2D shows plots beginning at the corresponding pump power. Square root dependence is shown in black solid line for reference. The corresponding theoretical treatment is in Fig.4 and the Supplementary Material.

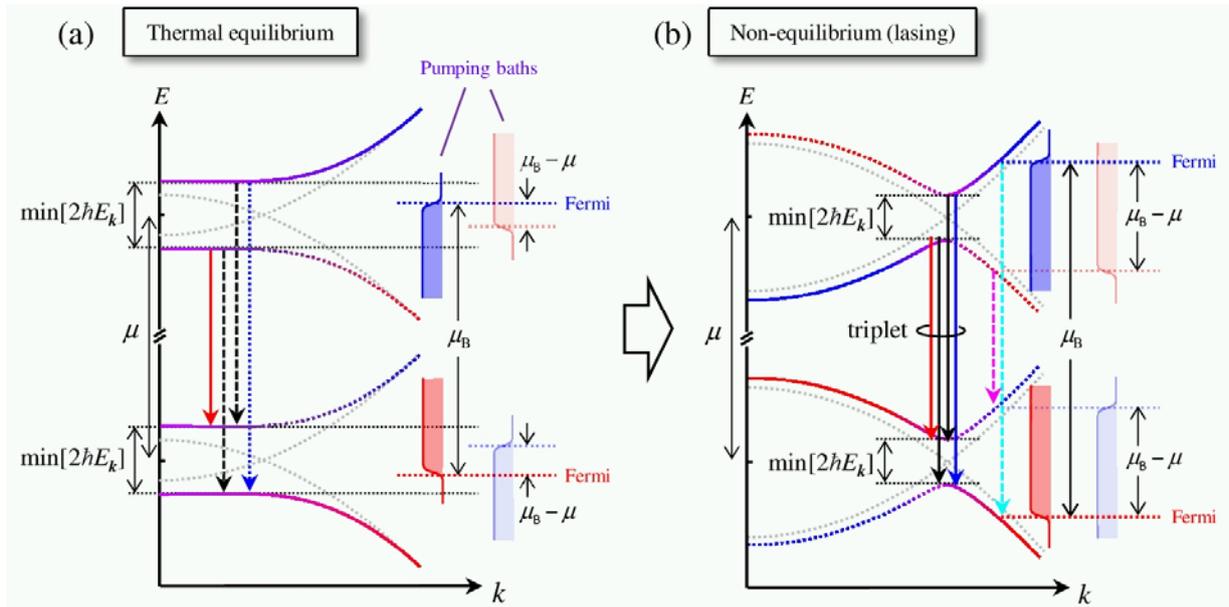

**Fig. 3|** Energy diagrams of the e-h-p system at thermal equilibrium (**A**) and non-equilibrium (**B**). Equilibrium (non-equilibrium) corresponding to low (high) excitation densities. In the low density regime, the high-energy peak of triplet is not visible since the excitation supplied from the pumping bath cannot exceed the gap, while it is visible at high density (panel **B**).

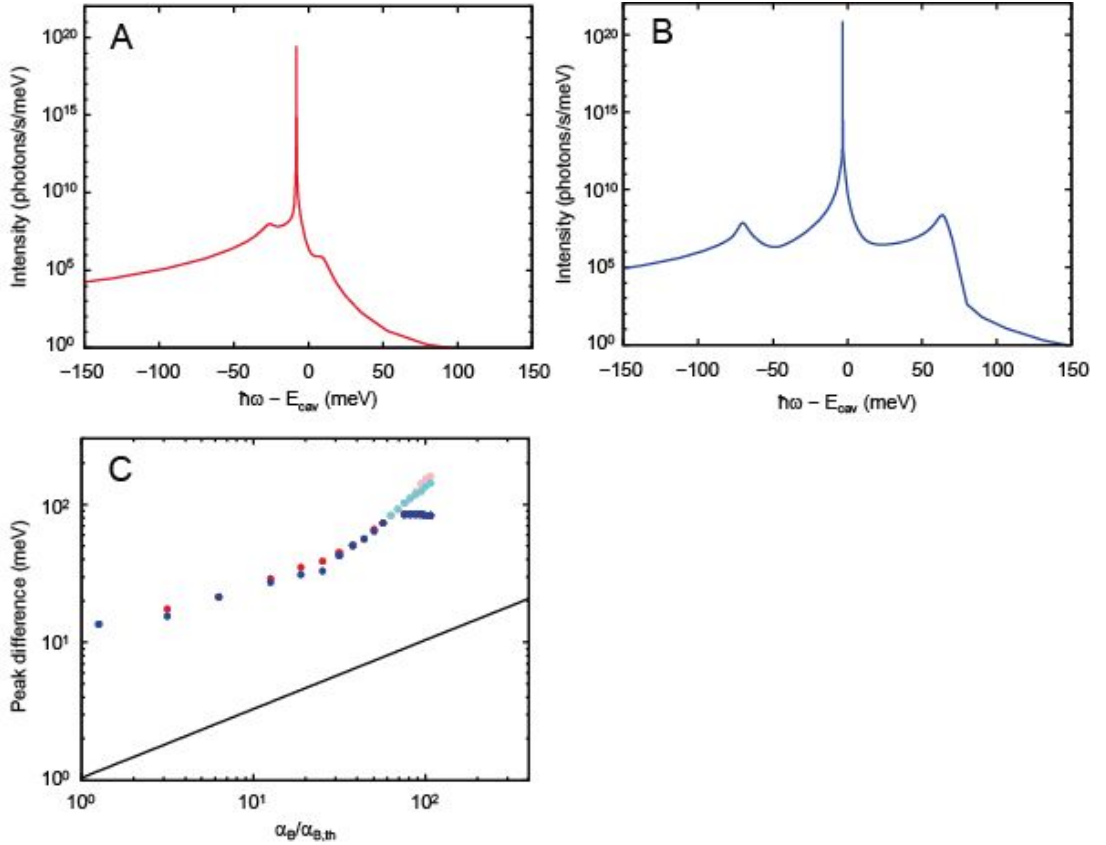

**Fig. 4**| Numerical results for the spectral properties. (A and B) Calculated photoluminescence spectrum $I_{SS}^{inc}(\omega, \mathbf{q}=0)$ at $\mu_B - E_{LP} = 5$ meV ((A), low excitation density) and 80 meV ((B), high excitation density) indicated by the arrows in Fig. 5B. The experimental energy separation between the two peaks are an order of magnitude smaller than the calculated values. This discrepancy may be caused by dephasing and polariton–polariton scattering processes not taken into account in our theory.

(C) Energy separation between the main peak and the side-band peaks as a function of the pumping strength $\alpha_B \equiv \mu_B - E_{LP}$. In order to compare the results with the experiments, $\alpha_B$ is again normalized by its threshold value $\alpha_{B,th} \equiv \mu_{B,th} - E_{LP}$ (= 1.64 meV). The blue (red) points denote the difference in energy positions between the higher-energy (lower-energy) sideband and the main peak, the transitions of which are indicated by the blue (red) and black arrows in Fig. 3. We note that, under the lasing conditions, the calculated spectra exhibit two additional weak peaks (not shown), corresponding to the aqua and pink transitions in Fig. 3 (B). We therefore plot the energy separation between the main peak and the two additional peaks by aqua and pink in panel (C). The PL intensities of the two additional peaks are smaller than that of the high-energy peak by blue. Therefore, the discussion about the existence of those peaks are not given in the text. The black solid line is a guide for the eye proportional to $(\alpha_B/\alpha_{B,th})^{1/2}$.

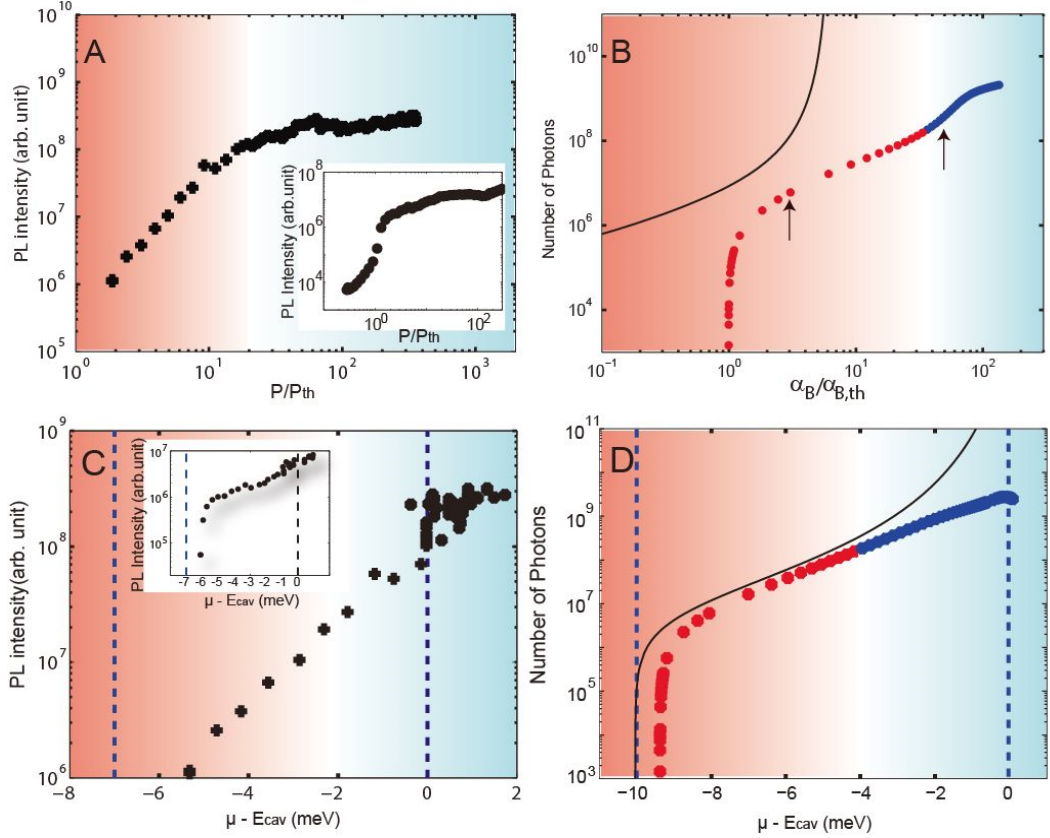

**Fig. 5|** Pump-power dependences of the PL intensities (**A** (experiment) and **B** (theory)) and the PL energy dependences of the main peak (**C** (experiment) and **D** (theory)). (**A**) PL intensity at the time of maximum PL where spectral integration is performed. The inset shows time-integrated data of peak PL intensity. (**B**) Number of photons inside the cavity as a function of the pumping level normalized by the threshold value $\alpha_{B,th} = \mu_{B,th} - E_{LP}$, for a cavity decay rate of $\hbar\kappa = 100$ μeV. The solid black curve corresponds to complete thermal equilibrium theory ($\kappa = 0$, $\gamma = 0^+$). The arrows in the panel indicate the numbers of photons for the photoluminescence spectra shown in Fig. 4A and 4B. (**C**) Experimental time-resolved data, focusing on the high-excitation regime, used to observe the behaviour around the cavity photon energy, where the spectral integration is performed. The inset shows the peak PL intensity from time-integrated measurements. (**D**) Calculated results. The vertical dashed lines of panel **C** and **D** are the LP and the cavity photon energy far below $P_{th}$. The red (blue) shaded areas (and also the dots in panels (**B** and **D**)) indicate that the system is at quasi-equilibrium (non-equilibrium). In the experiment (**A** and **C**), the crossover into non-equilibrium is judged by the appearance of the high-energy peak. However, since it is difficult to decide the crossover point clearly at a specific pump power due to the gradual appearance of the high-energy peak as shown in Fig. 2C, plots are given in black while gradual crossover is shown by shaded background from red to blue. The LP energy position of (C):-7 meV and (D) -10 meV are different since normal mode splitting of the experiment is 14 meV while that of theoretical calculation is 20 meV. However, the argument remains the same as long as the degree of normal splitting is on the same order. In other words, the qualitative feature is the same when the energy scale is normalized by the normal mode splitting.

# Supplementary Information
# High-energy side-peak emission of exciton-polariton condensates in high density regime

Tomoyuki Horikiri[1,2,3,4,5], Makoto Yamaguchi[4,6], Kenji Kamide[3,6], Yasuhiro Matsuo[1,3], Tim Byrnes[7,8,1], Natsuko Ishida[4], Andreas Löffler[9], Sven Höfling[9,10,1], Yutaka Shikano[11,12,13], Tetsuo Ogawa[6,14], Alfred Forchel[9], and Yoshihisa Yamamoto[1,2,3,4,15]

Correspondence to: yyamamoto@stanford.edu, horikiri@ynu.ac.jp

**This file includes:**
    Mollow triplet — similarities and differences
    Comparisons with a two-level model
    Theoretical treatments
    Figs. S1 to S9

## S1: Mollow triplet — similarities and differences

In quantum optics, it is well known that a spectral triplet can be observed in the fluorescence spectrum from a two-level atom driven by a resonant continuous-wave laser field. Such a triplet is called the Mollow triplet[23], and it has close relations to the spectra discussed in our study. Therefore, in this section, the similarities and differences between them are presented.

First, we briefly review the mechanism and the spectral properties of the Mollow triplet in atomic systems. When a laser field is applied to a single two-level atom, dressed states can be formed as a result of the strong coupling between the atom and photon states. Such dressed states are described by

$$|\pm, n\rangle = \frac{1}{\sqrt{2}}[|\text{excited}, n-1\rangle \pm |\text{ground}, n\rangle], \tag{S1}$$

where $n$ is the photon number and $|\text{ground}\rangle$ ($|\text{excited}\rangle$) is the ground (excited) state of the atom. The eigenenergies of these states are written as

$$E_{\pm,n} = n\hbar\omega_0 \pm \hbar\Omega_n/2, \tag{S2}$$

where $\omega_0$ is the laser frequency and $\Omega_n = 2g\sqrt{n+1}$ is the Rabi frequency. Therefore, in the strong-field limit ($n \gg 1$), $\Omega_n \approx \Omega_{n+1} \equiv \Omega_R$ is a good approximation, and four transitions become possible between the eigenstates (see Fig. S1A). As a result, three peaks are formed in the fluorescence spectrum. We note that the energy separation between the main peak and the sideband peaks, which corresponds to $\Omega_R$, is roughly proportional to the square root of the intensity of the incident laser (that is, it is roughly proportional to $\sqrt{n}$) in the strong-field limit. Such a fluorescence spectrum $S(\omega)$ can be written as

$$S(\omega) = \frac{I}{2\pi} \frac{\pi}{\gamma^2 + 2\Omega_R^2} \delta(\omega - \omega_0)$$
$$+ \frac{I}{2\pi} \left\{ \frac{1}{2} \frac{\gamma/2}{[\gamma/2]^2 + [\omega - \omega_0]^2} + \frac{1}{4} \frac{3\gamma/4}{[3\gamma/4]^2 + [\omega - [\omega_0 + \Omega_R]]^2} + \frac{1}{4} \frac{3\gamma/4}{[3\gamma/4]^2 + [\omega - [\omega_0 - \Omega_R]]^2} \right\},$$
$$\tag{S3}$$

where $I$ is a constant value and $\gamma$ is the relaxation rate of the atom. The first term is the elastic Rayleigh scattering of the laser field, and the last three terms correspond to the fluorescence spectrum. Here, the Rabi frequency $\Omega_R$ can be made larger than the order of $\gamma$ by using a sufficiently strong laser field. Hence, the spectra from the last three terms are well separated, resulting in the Mollow triplet, while the first elastic scattering term becomes negligible for $\Omega_R \gg \gamma$. From this expression, we can obtain the following spectral properties: the central peak at $\omega_0$ has a linewidth of $\gamma/2$, and the peak height is three times larger than the side peaks, while the side peaks of the frequencies $\omega_0 \pm \Omega_R$ have linewidths of $3\gamma/4$. In contrast, in the weak field limit

($\Omega_R \ll \gamma$), $S(\omega)$ shows a single-peak spectrum located at $\omega_0$. In what follows, this is referred to as a simple Mollow triplet to emphasize the conventional physical picture.

The simple Mollow triplet also appears in emission spectra when the atom is strongly coupled to a single-mode cavity[24]. In this case, the triplet can be found when a strong coherent light field is initially present inside the cavity. However, the vacuum Rabi splitting can be found when the coherent light field becomes sufficiently weak because the atom is still strongly coupled to the vacuum photon state of the cavity (see Fig. S1B). As a result, a smooth change from the vacuum Rabi splitting to the simple Mollow triplet can be seen in the emission spectra when the atom is strongly coupled to a cavity[24].

In the exciton–polariton case, Mollow-triplet-like spectra are also theoretically predicted, first in the equilibrium limits[1] and then in the non-equilibrium cases[19] where the side peaks are called amplitude modes and the excitons are simply taken as two-level systems. In the present work, the PL, not only in the equilibrium case but also in the non-equilibrium case, has been calculated by explicitly treating the semiconductor electrons and holes with their Coulomb interactions (see also Section 3 for the calculation methods). The spectral features in this case are not completely the same as those of the simple Mollow triplet because of, e.g., the dispersions of electrons and holes and the many-body effects[31,32], but the physical picture analogous to the Mollow triplet is still useful for understanding their emission mechanisms. The typical energy diagrams in a high-excitation regime are shown in Fig. S2. The energy dispersions of the electrons and holes undergo the band-gap renormalization (BGR) when only the electron–electron and hole–hole Coulomb interactions are considered (Fig. S2A). Then, the energy gap of $\min[2\hbar E_k] \sim 2\hbar\Delta$ appears in each band, as shown in Fig. S2B, if we also switch on the dipole coupling to the cavity photons and the electron–hole (e–h) attractive Coulomb interaction. The formal definitions of the minimum e–h pair breaking energy $\min[2\hbar E_k]$ and the generalized Rabi frequency $\Delta_k$ can be found in Subsection S3.2. Here, the energy gap is formed not only by the standard Rabi splitting due to the strong light field inside the cavity[24] but also by the BCS-like e–h Coulomb correlations[22,26-28]. Thus, the spectral triplet can also be predicted in the semiconductor model.

However, unlike the simple Mollow triplet, the intensity ratio of the side peaks can change, depending on the steady-state conditions. In the case of thermal equilibrium, the energy needed to break e–h pairs, $\min[2\hbar E_k]$, is larger than $\mu_B - \mu + 2\hbar\gamma + 2k_B T$ (see Ref. 22 and also Subsection S3.2), a typical example of which is shown in Fig. S3A. Here, $\mu_B$, $\mu$, $1/\gamma$, and $T$ represent the chemical potential of the electron and hole pumping baths (which we refer to as the pumping parameter in the main text, since it measures the degree of pumping), the oscillation frequency of the photon and polarization fields, the thermalization time scale of the pumping baths, and the temperature defined in the pumping baths. In this case, the Fermi surface of the pumping baths cannot go beyond the energy gap. As a result, the lower-energy peak (indicated by the solid red arrow) becomes large because sufficient electrons and holes can be obtained from the pumping baths to recombine and emit photons into the low-energy peak. However, the situation is opposite for the high-energy peak (the dotted blue line); electrons and holes are not supplied by the pumping baths. Hence, the low-energy peak becomes larger than the high-energy peak in the thermal equilibrium regimes. In contrast, non-equilibrium effects cannot be negligible when $\mu_B - \mu$ becomes larger than $\min[2\hbar E_k] - 2\hbar\gamma - 2k_BT$, and there are ***k***-regions described by the Maxwell–semiconductor–Bloch equations (MSBE) used in semiconductor lasers when $\mu_B - \mu \gtrsim \min[2\hbar E_k] + 2\hbar\gamma + 2k_BT$ (see also Subsection S3.2). In this case, the Fermi surface of the pumping baths exceeds the energy gap $\min[2\hbar E_k]$, a typical example of which is shown in Fig. S3B. Hence, the high-energy peak can become stronger than in the thermal equilibrium cases, which is significant since thermal equilibrium can cause fairly strong high-energy peaks.

Furthermore, to understand the emission mechanism of exciton–polariton microcavity systems, the important role that many-body effects play must also be considered. In semiconductor materials (with no cavity), it is well known that e–h Coulomb interactions cause significant enhancement of the PL intensity around the Fermi energy even for a plasma state[31,32]. This effect also should be treated in our case, and as a result, the high-energy peak is further enhanced when the Fermi energy (which is given by $\mu_B$) roughly coincides with the high-energy peak located at $\mu + \min[2\hbar E_k]$. In fact, in our calculations, the intensity of the high-energy peak exceeds that of the low-energy peak only when $\mu_B - \mu \sim \min[2\hbar E_k]$. In this sense, it is clear that the many-body effects also play an essential role, which is absent in, e.g., the two-level model (see also Section S2). It is instructive to note that the condition $\mu_B - \mu \sim \min[2\hbar E_k]$ is equivalent to the crossover condition from quasi-equilibrium phases ($\mu_B - \mu \lesssim \min[2\hbar E_k]$) into lasing phases ($\mu_B - \mu \gtrsim \min[2\hbar E_k]$) when ignoring the effects of $\gamma$ and $T$ [16]. Therefore, we can conclude that the system is near the crossover regime into lasing if the high-energy peak is stronger than the low-energy peak, at least in principle.

## S2: Comparisons with a two-level model

In the present work, the theoretical discussions are based on the framework explained in the next section (Section S3), where electrons and holes are explicitly treated with their Coulomb interactions and the many-body effects resulting from the Coulomb interactions are included within the Hartree–Fock (HF) level. However, it is important to compare the results with another model where the e–h system is simply treated as a set of two-level systems (TLSs)[19,43,44]. Hence, in this section, we present numerical results based on such a TLS model and show that the height of the high-energy peak does not exceed that of the low-energy peak. We note that some results include new viewpoints on the result that the quasi-equilibrium and non-equilibrium regimes are clearly distinguished even for the TLS model.

The model discussed here is schematically shown in Fig. S5 with the $n$ two-level systems. In our calculations, the cavity is resonant with the transition energy $\hbar\omega_x$, and the lower polariton energy $E_{LP}$ is located $\hbar g\sqrt{n} = 10$ meV below the bare cavity, where the coupling constant between the two-level atom and the cavity mode are assumed to be $\hbar g = 1.0$ meV and $n = 100$, respectively. In addition, $T = 10$ K, $\hbar\kappa = 100$ μeV, and $\hbar\gamma = 4$ meV are used, which are the same as the parameters used in the main text (see also Subsection S3.5) with the decay rate from the cavity mode $\kappa$.

Figure S6A shows the calculated photon number inside the cavity (= $|a_0|^2$), in the same manner as Fig. 2B in the main text. The number of photons calculated with the thermal equilibrium theory (black solid line: $\kappa = 0$, $\gamma = 0^+$) diverges when $\mu_B$ approaches the bare cavity energy $E_{cav}$ for the same reasons as in Fig. 2B. In contrast, the number of photons does not diverge for the non-zero $\kappa$ and $\gamma$. Here, red data points and lines are used in the plots when the system can still be regarded as in a quasi-equilibrium state, and otherwise blue plot elements are used. The typical emission spectra for quasi-equilibrium and non-equilibrium regimes are shown in Fig. S6, B and C. In the case of quasi-equilibrium, the low-energy peak is stronger than the high-energy peak, which is qualitatively the same as is shown in Fig. 2C. Then, the high-energy peak becomes stronger when the system enters the non-equilibrium regime (Fig. S6C). However, the high-energy peak does not exceed the low-energy peak, which is in stark contrast to the semiconductor model where the Coulomb interactions are explicitly taken into account. As described in Section S1, BCS-like Coulomb correlations also play an important role in the intensity of the high-energy peak in the semiconductor model. However, such a mechanism does not work in the TLS model because there is no Coulomb interaction. As a result, the high-energy peak becomes stronger than in the quasi-equilibrium cases but cannot exceed the low-energy peak in the TLS model. This clearly demonstrates that the many-body effects resulting from Coulomb interactions are indispensable for understanding why the high-energy peak is stronger than the low-energy one. Thus, the results from the TLS model also support the idea that the many-body effects have a strong influence on the intensity of the high-energy peak.

## S3: Theoretical treatments

In our study, we have employed the recently developed formalism described in Ref. 11 in order to investigate the equilibrium and non-equilibrium natures of the polariton condensate in high-density regimes (for a brief introduction, see also Ref. 17). In this treatment, a closed set of equations for the polarization function $p_k$, the number of electrons $n_{e,k}$, the number of holes $n_{h,k}$, and the cavity photon field $a_0$ is analyzed for the steady state. The advantage of this approach is the applicability to high-density regimes because electrons and holes are explicitly treated with their Coulomb interactions. This is in contrast to the other approaches based on the non-equilibrium Gross–Pitaevskii (GP) equation. Although the GP equation has many good points and enables intuitive and simple analyses[45,46], the exciton–polaritons are just taken as interacting bosons, which is a model that is only applicable to the low-density regimes. As a result, the fermionic features of the electrons and holes are completely neglected. On the other hand, our approach can be used to study the BEC–BCS–LASER crossover without any limit on the density. Furthermore, the conditions for the system to be well described by quasi-equilibrium theories can also be obtained explicitly. Hence, the formalism is quite helpful for the present work. However, how to discuss the emission spectra from the cavity is not presented in Ref. 22. Therefore, first, the formalism described in Ref. 22 is briefly reviewed in Subsections S3.1 and S3.2, and then, Subsections S3.3 and S3.4 explain how to calculate the emission spectra, which is an extension of the work for two-level systems[19,433,444]. Finally, the numerical procedures and parameters for our calculations are summarized in Subsection S3.5.

S3.1: Model and Hamiltonians

Let us recapitulate the relevant model and Hamiltonians. The model of the electron–hole–photon (e–h–p) system is shown in Fig. S7, where the reservoir responsible for the photonic leakage is the free-space vacuum fields (photon bath) and those for pumping are the pumping baths, which inject (extract) electrons and holes into (from) the e–h system on a time scale of $1/\gamma$ by emulating the thermalization processes of the e–h system.

The total Hamiltonian in the model is $\hat{H} = \hat{H}_{S} + \hat{H}_{R} + \hat{H}_{SR}$, where

$$\hat{H}_{S} = \hat{H}_{\text{kin}} + \hat{H}_{\text{el-el}} + \hat{H}_{\text{el-ph}}, \tag{S4}$$

$$\hat{H}_{R} = \hbar \sum_{\alpha=c,v} \sum_{k} \xi_{\alpha,k}^{B} \hat{b}_{\alpha,k}^{\dagger} \hat{b}_{\alpha,k} + \hbar \sum_{p} \xi_{p}^{B} \hat{\Psi}_{p}^{\dagger} \hat{\Psi}_{p}, \tag{S5}$$

$$\hat{H}_{SR} = \hbar \sum_{\alpha=c,v} \sum_{k,q} \Gamma_{k}^{\alpha} \left( \hat{c}_{\alpha,k}^{\dagger} \hat{b}_{\alpha,q} + \text{H.c.} \right) + \hbar \sum_{q,p} \zeta_{q} \left( \hat{a}_{q}^{\dagger} \hat{\Psi}_{p} + \text{H.c.} \right), \tag{S6}$$

are the system and reservoir Hamiltonians and their interaction Hamiltonian, respectively. Here, $\hat{c}_{c,k}$ and $\hat{c}_{v,k}$ denote annihilation operators for electrons with wavenumber $\mathbf{k}$ in the conduction (c) and valence (v) bands, respectively, and $\hat{a}_q$ is an annihilation operator for photons with wavenumber $\mathbf{q}$ in the cavity. Similarly, $\hat{b}_{c,k}$ and $\hat{b}_{v,k}$ denote fermion annihilation operators of pumping baths, and $\hat{\Psi}_p$ is the boson annihilation operator of free-space vacuum fields. Here, $\xi^B_{\alpha,k}$ and $\xi^B_p$ in Eq. (S5) are defined as

$$\xi^B_{c/v,k} \equiv \omega^B_{c/v,k} \mp \hbar^{-1}\mu/2, \quad \xi^B_p \equiv \omega^B_{ph,q} - \hbar^{-1}\mu, \tag{S7}$$

where $\hbar^{-1}\mu$ is the oscillation frequency of the coherent photon field[22], which will be determined later. $\Gamma^\alpha_k$ and $\zeta_k$ in Eq. (S6) are the coupling constants between the system and each reservoir, satisfying the relations

$$\pi|\Gamma^\alpha_k|^2 D^B_\alpha(\omega) \cong \gamma_{\alpha,k} \cong \gamma, \quad \pi|\zeta_k|^2 D^B_{ph}(\omega) \cong \kappa_k \cong \kappa, \tag{S8}$$

with the following definitions of the density of states:

$$D^B_\alpha(\omega) \equiv \sum_k \delta(\xi^B_{\alpha,k} - \omega), \quad D^B_{ph}(\omega) \equiv \sum_p \delta(\xi^B_p - \omega). \tag{S9}$$

Here, the dependence on the wavenumber is neglected in Eq. (S8) for simplicity. $\hat{H}_{kin}$, $\hat{H}_{el\text{-}el}$, and $\hat{H}_{el\text{-}ph}$ in Eq. (S4) are described as

$$\hat{H}_{kin} = \hbar \sum_{\alpha,k} \xi_{\alpha,k} \hat{c}^\dagger_{\alpha,k} \hat{c}_{\alpha,k} + \hbar \sum_q \xi_{ph,q} \hat{a}^\dagger_q \hat{a}_q, \tag{S10}$$

$$\hat{H}_{el\text{-}el} = \frac{1}{2} \sum_{k,k',q} \sum_{\alpha,\alpha'} U'_q \hat{c}^\dagger_{\alpha,k+q} \hat{c}^\dagger_{\alpha',k'-q} \hat{c}_{\alpha',k'} \hat{c}_{\alpha,k}, \tag{S11}$$

$$\hat{H}_{el\text{-}ph} = -\hbar \sum_{k,q} \left( g^* \hat{a}_q \hat{c}^\dagger_{c,k+q} \hat{c}_{v,k} + g \hat{c}^\dagger_{v,k} \hat{c}_{c,k+q} \hat{a}^\dagger_q \right), \tag{S12}$$

where

$$\xi_{c/v,k} \equiv \omega_{c/v,k} \mp \hbar^{-1}\mu/2, \quad \xi_{ph,q} \equiv \omega_{ph,q} - \hbar^{-1}\mu, \tag{S13}$$

which are similar to the definitions in Eq. (S7). Here, $\omega_{c,k}$, $\omega_{v,k}$, and $\omega_{ph,q}$ denote the conduction-band, valence-band, and photonic dispersion relationships, respectively. In addition, $U'_q$ denotes the Coulomb interaction defined as

$$U'_q \equiv \begin{cases} U_q & \text{for } \mathbf{q} \neq 0 \\ 0 & \text{for } \mathbf{q} = 0 \end{cases}. \tag{S14}$$

We note that the Hamiltonian presented here represents the frequency-shifted picture in Ref. 22, but the shifts (Eqs. (S7) and (S13)) are made slightly different from the original one in order to obtain the final expression in a symmetric form.

S3.2: A closed set of mean-field equations for the BEC–BCS-LASER crossover
By using the Hamiltonians presented above, within the HF approximation, the standard Green's function and Heisenberg–Langevin approach yield the closed set of equations (1)-(7) in the main text. These equations are the same as obtained in Ref. 22 even though the expressions appear different[17]. In the derivation, it is assumed that a coherent photon field can be formed only in the $\mathbf{q} = 0$ state

$$\langle \hat{a}_q \rangle = \delta_{q,0} \langle \hat{a}_0 \rangle, \tag{S15}$$

and that the system is in a steady state described by the polarization function $p_k$, the number of electrons $n_{e,k}$, the number of holes $n_{h,k}$, and the cavity photon field $a_0$, respectively, which are defined as

$$p_k \equiv \langle \hat{c}^\dagger_{v,k} \hat{c}_{c,k} \rangle, \quad n_{e,k} \equiv \langle \hat{c}^\dagger_{c,k} \hat{c}_{c,k} \rangle, \quad n_{h,k} \equiv 1 - \langle \hat{c}^\dagger_{v,k} \hat{c}_{v,k} \rangle, \quad \langle \hat{a}_0 \rangle = a_0. \tag{S16}$$

In Eqs. (1)–(7), the notation is transformed into the e–h picture

$$\omega_{e,k} \equiv \omega_{c,k}, \quad \omega_{h,k} \equiv -\omega_{v,k} + \hbar^{-1} \sum_q U'_{q-k}, \tag{S17}$$

with the Coulomb-renormalized dispersion relation

$$\tilde{\omega}_{\alpha,k} \equiv \omega_{\alpha,k} + \Sigma^{\text{BGR}}_{\alpha,k}, \quad \hbar \Sigma^{\text{BGR}}_{\alpha,k} \equiv -\sum_q U'_{q-k} n_{\alpha,q}, \tag{S18}$$

where $\alpha \in \{e, h\}$. Here, the unknown variables for the closed set of equations (1)–(7) are $a_0$, $p_k$, $n_{e,k}$, $n_{h,k}$, and $\mu$. Thus, we have shown a set of coupled equations. Eqs. (1)-(7), then, yields

$$\Delta_k \cong \sum_q U^{\text{eff},\kappa}_{q,k} \Delta_q \frac{\tanh(\beta \hbar E_q / 2)}{2\hbar E_q}, \tag{S19}$$

$$U^{\text{eff},\kappa}_{q,k} \equiv \frac{\hbar |g|^2}{\omega_{\text{ph},0} - \hbar^{-1}\mu - i\kappa} + U'_{q-k}. \tag{S20}$$

when $\min[2\hbar E_k] \gtrsim \mu_B - \mu + 2\hbar\gamma + 2k_B T$ by assuming $\omega_{e,k} = \omega_{h,k}$ and $\mu_e^B = \mu_h^B = \mu_B/2$ for simplicity[22,16,17]. In this way, Eq. (S19) is reduced to the BCS gap equation for electrons and holes used in thermal equilibrium theories[20,21]. Therefore, in this situation, $\mu$ and $\min[2\hbar E_k]$ denote the chemical potential of the e–h–p system and the minimum energy required for breaking an e–h pair, respectively. Thus, the system can be well described by quasi-thermal equilibrium theories when the condition $\min[2\hbar E_k] \gtrsim \mu_B - \mu + 2\hbar\gamma + 2k_B T$ is satisfied. This situation is already shown in Fig. S3A.

In contrast, non-equilibrium effects cannot be neglected for $\mu_B - \mu \gtrsim \min[2\hbar E_k] - 2\hbar\gamma - 2k_B T$. Furthermore, Eqs. (1)–(7) result in

$$0 = -i\xi_{\text{ph},0} a_0 + ig \sum_k p_k - \kappa a_0, \tag{S21}$$

$$0 = -i[\tilde{\xi}_{e,k} + \tilde{\xi}_{h,k}] p_k - i\Delta_k N_k - 2\gamma p_k, \tag{S22}$$

$$0 = -2\,\text{Im}[\Delta_k p^*_k] - 2\gamma[n_{e(h),k} - f_{e(h),k}], \tag{S23}$$

when $\mu_B - \mu \gtrsim 2\hbar E_k + 2\hbar\gamma + 2k_B T$ (which is $k$-dependent) is satisfied. Again, for simplicity, it is assumed that $\omega_{e,k} = \omega_{h,k}$ and $\mu_e^B = \mu_h^B$. Here, $N_k \equiv n_{e,k} + n_{h,k} - 1$ is the population inversion of the system and $f_{e(h),k} \equiv [\exp\{\beta(\hbar\tilde{\omega}_{e(h),k} - \mu^B_{e(h)})\} + 1]^{-1}$ is the Fermi distribution function. Equations (S21)–(S23) are the very MSBE under the RTA used for steady-state semiconductor lasers[466]. Therefore, in this situation, $\hbar^{-1}\mu$ and $\Delta_k$ denote the laser oscillation frequency and the generalized Rabi frequency, respectively[26]. Thus, the $k$-dependent condition $\mu_B - \mu \gtrsim 2\hbar E_k + 2\hbar\gamma + 2k_B T$ determines the *k*-region where use of the MSBE is justified. In other words, there are *k*-regions described by the MSBE when $\mu_B - \mu \gtrsim \min[2\hbar E_k] + 2\hbar\gamma + 2k_B T$, and an example of this situation is shown in Fig. S3B. Here, we note that $\Delta_k \cong g^*\sqrt{n} \cong 2\Omega_R$ ($a_0 \approx \sqrt{n}$) causes the exact Rabi splitting if there is no Coulomb term in the definition of $\Delta_k \equiv g^* a_0 + \hbar^{-1}\sum_q U'_{q-k} p_q$, but $\Delta_k$ is also influenced by the BCS-like e–h Coulomb correlation term $\sum_q U'_{q-k} p_q$. In this sense, the energy gap (Fig. S2B and Fig. S3B) is formed not only by the standard Rabi splitting due to the strong field but also by the BCS-like e–h Coulomb correlations.

S3.3: Photoluminescence spectra

Based on standard quantum optics, the steady-state emission spectra observed outside the cavity can be described as

$$I_{SS}(\omega, \boldsymbol{q}) = \frac{\kappa}{\pi} \lim_{t \to \infty} \int_{-\infty}^{\infty} d\tau \langle \hat{a}_q^\dagger(t) \hat{a}_q(t+\tau) \rangle e^{i\omega\tau}. \tag{S24}$$

The emission spectra can then be divided into coherent and incoherent parts

$$I_{SS}(\omega, \boldsymbol{q}) = I_{SS}^{coh}(\omega, \boldsymbol{q}) + I_{SS}^{inc}(\omega, \boldsymbol{q}), \tag{S25}$$

with definitions of

$$I_{SS}^{coh}(\omega, \boldsymbol{q}) \equiv 2\kappa |a_0|^2 \delta_{q,0} \delta(\omega), \tag{S26}$$

$$I_{SS}^{inc}(\omega, \boldsymbol{q}) \equiv \frac{\kappa}{\pi} \lim_{t \to \infty} \int_{-\infty}^{\infty} d\tau \langle \Delta\hat{a}_q^\dagger(t) \Delta\hat{a}_q(t+\tau) \rangle e^{i\omega\tau}, \tag{S27}$$

where $\Delta\hat{a}_q(t) \equiv \hat{a}_q(t) - \langle \hat{a}_q(t) \rangle$. Hence, the incoherent part of the emission spectra can be described by the photon Green's functions in the frequency domain as

$$I_{SS}^{inc}(\omega, \boldsymbol{q}) = i\hbar \frac{\kappa}{2\pi} \left( D^K(\omega;11;\boldsymbol{q}) - D^R(\omega;11;\boldsymbol{q}) + D^A(\omega;11;\boldsymbol{q}) \right), \tag{S28}$$

where $D^R$, $D^A$, and $D^K$ are respectively the retarded, advanced, and Keldysh parts of the closed-time path of the (non-condensed) photon Green's function defined as

$$D_C(\tau_1\tau_2; \sigma_1\sigma_2; \boldsymbol{q}_1\boldsymbol{q}_2) \equiv \frac{1}{i\hbar} \langle T_C[\Delta\hat{\psi}_{\sigma_1,q_1}(\tau_1) \Delta\hat{\psi}_{\sigma_2,q_2}^\dagger(\tau_2)] \rangle, \tag{S29}$$

where $T_C$ is the time-ordering operator on the closed-time path[47,48]; $\sigma_1, \sigma_2 \in \{1, 2\}$; and the Heisenberg operators are

$$\Delta\hat{\psi}_{1,q}(\tau) \equiv \hat{a}_q(\tau) - \langle \hat{a}_q(\tau) \rangle, \quad \Delta\hat{\psi}_{2,q}(\tau) \equiv \hat{a}_{-q}^\dagger(\tau) - \langle \hat{a}_{-q}^\dagger(\tau) \rangle. \tag{S30}$$

In this context, the emission spectra can be discussed when $D^R$, $D^A$, and $D^K$ in the frequency domain are obtained. Here, $D^R$, $D^A$, and $D^K$ can be calculated if the self-energy $\Sigma^{ph}(\omega; \boldsymbol{q})$ for the photon Green's function is determined because the Dyson equation for the photon Green's function is described as

$$D(\omega; \boldsymbol{q}) = [[D^0(\omega; \boldsymbol{q})]^{-1} - \Sigma^{ph}(\omega; \boldsymbol{q})]^{-1}, \tag{S31}$$

in the matrix form of

$$X(\omega;\boldsymbol{q}) \equiv \begin{pmatrix} X^{\text{R}}(\omega;11;\boldsymbol{q}) & X^{\text{R}}(\omega;12;\boldsymbol{q}) & X^{\text{K}}(\omega;11;\boldsymbol{q}) & X^{\text{K}}(\omega;12;\boldsymbol{q}) \\ X^{\text{R}}(\omega;21;\boldsymbol{q}) & X^{\text{R}}(\omega;22;\boldsymbol{q}) & X^{\text{K}}(\omega;21;\boldsymbol{q}) & X^{\text{K}}(\omega;22;\boldsymbol{q}) \\ 0 & 0 & X^{\text{A}}(\omega;11;\boldsymbol{q}) & X^{\text{A}}(\omega;12;\boldsymbol{q}) \\ 0 & 0 & X^{\text{A}}(\omega;21;\boldsymbol{q}) & X^{\text{A}}(\omega;22;\boldsymbol{q}) \end{pmatrix}, \quad (S32)$$

where $D^0$, $D$, and $\Sigma^{\text{ph}}$ are described by $X$ for notational simplicity. Therefore, in the following, we show the self-energy used for calculating $D^{\text{R}}$, $D^{\text{A}}$, $D^{\text{K}}$ (Eq. (S31)), and $I_{\text{SS}}^{\text{inc}}(\omega,\mathbf{q})$ (Eq. (S27)).

S3.4: Self-energies for the photon Green's function

Photons in the cavity are basically influenced by the e–h system and by the free-space vacuum fields (photon bath), as shown in Fig. S5 and in the Hamiltonians in Subsection S3.1. As a result, the self-energy for the photon Green's function in our study is written as

$$\Sigma^{\text{ph}}(\omega;\boldsymbol{q}) = \Sigma_{\text{SR}}^{\text{ph}}(\omega;\boldsymbol{q}) + \Pi(\omega;\boldsymbol{q}), \quad (S33)$$

$\Sigma_{\text{SR}}^{\text{ph}}(\omega;\boldsymbol{q})$ and $\Pi(\omega;\boldsymbol{q})$ are the self-energies due to the system–reservoir coupling (Eq. (S6)) and the light–matter coupling (Eq. (S12)), respectively. The latter describes the carrier-induced index change in conventional semiconductor laser theory[26]. By assuming that the photon-bath states are vacuum states, the self-energy $\Sigma_{\text{SR}}^{\text{ph}}$ can be written in the matrix form (Eq. (S32)) as

$$\Sigma_{\text{SR}}^{\text{ph}}(\omega;\boldsymbol{q}) = \begin{pmatrix} -i\hbar\kappa & 0 & -i2\hbar\kappa & 0 \\ 0 & +i\hbar\kappa & 0 & -i2\hbar\kappa \\ 0 & 0 & +i\hbar\kappa & 0 \\ 0 & 0 & 0 & -i\hbar\kappa \end{pmatrix} \quad (S34)$$

within the approximation of Eq. (S8) (see also Fig. S8A). In contrast, the self-energy $\Pi$ can be written as

$$\Pi(\nu;\boldsymbol{q}) = -[i\hbar]^3 g^2 \sum_{\boldsymbol{k}_1,\boldsymbol{k}_2} \widetilde{K}_{0,\boldsymbol{q}}(\nu;\boldsymbol{k}_1\boldsymbol{k}_2), \quad (S35)$$

where $\widetilde{K}_{0,\boldsymbol{q}}(\nu;\boldsymbol{k}_1\boldsymbol{k}_2)$ is a two-particle Green's function. In Eq. (S35), $g$ is assumed to be real for simplicity, and $\widetilde{K}_{0,\boldsymbol{q}}(\nu;\boldsymbol{k}_1\boldsymbol{k}_2)$ takes the same matrix form as Eq. (S32). Thus, it turns out that the two-particle Green's function $\widetilde{K}_{0,\boldsymbol{q}}(\nu;\boldsymbol{k}_1\boldsymbol{k}_2)$ is required to calculate the self-energy $\Pi(\omega;\boldsymbol{q})$. However, in general, it is difficult to accurately obtain the two-particle Green's function when the effects of Coulomb interactions are present. Therefore, in this study, the $T$-matrix approximation is applied, as shown in Fig. S8,

$$\widetilde{K}_{0,\boldsymbol{q}}(\omega;\boldsymbol{k}_1\boldsymbol{k}_2) = -K_{0,\boldsymbol{q}}(\omega;\boldsymbol{k}_1\boldsymbol{k}_2) - K_{0,\boldsymbol{q}}(\omega;\boldsymbol{k}_1\boldsymbol{k}_1) T(\omega;\boldsymbol{k}_1\boldsymbol{k}_2) K_{0,\boldsymbol{q}}(\omega;\boldsymbol{k}_4\boldsymbol{k}_2), \quad (S36)$$

$$T(\omega;\boldsymbol{k}_1\boldsymbol{k}_2) = i\hbar U'_{\boldsymbol{k}_1-\boldsymbol{k}_2} \boldsymbol{I}_{4\times 4} + i\hbar \sum_{\boldsymbol{k}_3,\boldsymbol{k}_4} U'_{\boldsymbol{k}_1-\boldsymbol{k}_3} K_{0,\boldsymbol{q}}(\omega;\boldsymbol{k}_3\boldsymbol{k}_4) T(\omega;\boldsymbol{k}_4\boldsymbol{k}_2), \quad (S37)$$

where $I_{4\times 4}$ is the unit 4×4 matrix and $K_{0,q}(\omega;k_1k_2)$ is the two-particle Green's function without any interactions. Therefore, in Eqs. (S36) and (S37), the matrix elements of $K_{0,q}(\omega;k_1k_2)$ can be described by using the one-particle Green's functions

$$K_{0,q}^{R/A}(\omega;\alpha_1\alpha_2;k_1k_2) = \frac{\delta_{k_1,k_2}}{2} \int_{-\infty}^{\infty} \frac{d\nu}{2\pi} \left\{ \begin{array}{l} G^K(\nu-\omega;\bar{\alpha}_2\bar{\alpha}_1;k_1-q/2)G^{R/A}(\nu;\alpha_1\alpha_2;k_1+q/2) \\ + G^{A/R}(\nu-\omega;\bar{\alpha}_2\bar{\alpha}_1;k_1-q/2)G^K(\nu;\alpha_1\alpha_2;k_1+q/2) \end{array} \right\},$$
(S38)

$$K_{0,q}^{K}(\omega;\alpha_1\alpha_2;k_1k_2) = \frac{\delta_{k_1,k_2}}{2} \int_{-\infty}^{\infty} \frac{d\nu}{2\pi} \left\{ \begin{array}{l} G^K(\nu-\omega;\bar{\alpha}_2\bar{\alpha}_1;k_1-q/2)G^K(\nu;\alpha_1\alpha_2;k_1+q/2) \\ + G^R(\nu-\omega;\bar{\alpha}_2\bar{\alpha}_1;k_1-q/2)G^A(\nu;\alpha_1\alpha_2;k_1+q/2) \\ + G^A(\nu-\omega;\bar{\alpha}_2\bar{\alpha}_1;k_1-q/2)G^R(\nu;\alpha_1\alpha_2;k_1+q/2) \end{array} \right\},$$
(S39)

where $\alpha_1, \alpha_2 \in \{1, 2\}$, $\bar{\alpha}_i = 2, 1$ for $\alpha_i = 1, 2$, and the one-particle Green's functions can be written as

$$G_k^{R/A}(\nu) = \begin{pmatrix} \hbar\nu - \hbar\tilde{\xi}_{e,k} \pm i\hbar\gamma & \hbar\Delta_k \\ \hbar\Delta_k^* & \hbar\nu + \hbar\tilde{\xi}_{h,k} \pm i\hbar\gamma \end{pmatrix}^{-1},$$
(S40)

$$G^K(\nu;k) = -i2\hbar\gamma \cdot G^R(\nu;k) \begin{pmatrix} 1-2f_e^B(\nu) & 0 \\ 0 & 2f_h^B(-\nu)-1 \end{pmatrix} G^A(\nu;k),$$
(S41)

in the matrix form of

$$G^{R/A/K}(\nu;k) \equiv \begin{pmatrix} G^{R/A/K}(\nu;11;k) & G^{R/A/K}(\nu;12;k) \\ G^{R/A/K}(\nu;21;k) & G^{R/A/K}(\nu;22;k) \end{pmatrix}.$$
(S42)

These Green's functions (Eqs. (S40) and (S41)) were already obtained from the solutions of Eqs. (1)–(7). As a result, the $T$ matrix and the two-particle Green's function can be calculated with Eqs. (S36) and (S37). Hence, the self-energy $\Pi(\omega;q)$ can be calculated. Thus, at this stage, the formulations for the emission spectra are complete. For readers unfamiliar with Green's functions, the procedures for calculating the emission spectra are given in Subsection S3.5, where the parameters used in our study are also described.

Section S3.5: Procedures and parameters for calculating emission spectra

The formalism for calculating the emission spectra has been presented in Subsections S3.3 and S3.4. The procedures for calculating the emission spectra are now summarized as follows:

**Step 1**: Solve the simultaneous system of equations consisting of Eqs. (1)–(7) for the unknown variables $a_0$, $p_k$, $n_{e,k}$, $n_{h,k}$, and $\mu$.

**Step 2**: Evaluate the one-particle Green's functions (Eqs. (S40) and (S41)) by using the values of $a_0$, $p_k$, $n_{e,k}$, $n_{h,k}$, and $\mu$ obtained in Step 1.

**Step 3**: Evaluate the two-particle Green's function $K_{0,q}(\omega;k_1k_2)$ (Eqs. (S38) and (S39)) by using the one-particle Green's functions obtained in Step 2.

**Step 4**: Find the *T* matrix satisfying Eq. (S37) for $K_{0,q}(\omega;k_1k_2)$ obtained in Step 3.

**Step 5**: Calculate the two-particle Green's function $\tilde{K}_{0,q}(\omega;k_1k_2)$ in Eq. (S36) by using the *T* matrix obtained in Step 4.

**Step 6**: Calculate the self-energy $\Sigma^{\text{ph}}(\omega; q)$ for the photon Green's function (Eq. (S33) with Eqs. (S34) and (S35)) by using $\tilde{K}_{0,q}(\omega;k_1k_2)$ obtained in Step 5.

**Step 7**: Calculate the photon Green's function $D(\omega; q)$ by using Eq. (S31) and the value of $\Sigma^{\text{ph}}(\omega; q)$ obtained in Step 6.

**Step 8**: Calculate the emission spectra $I_{\text{SS}}^{\text{inc}}(\omega,q)$ by using the photon Green's function obtained in Step 7.

By following Steps 1−8, the spectra $I_{\text{SS}}^{\text{inc}}(\omega,q=0)$ in Fig. 5 in the main text were obtained. In our calculations, the contact potential model $U_q = U$ is used for simplicity, and therefore, the *k* dependence of $\Delta_k$ is eliminated as $\Delta_k = \Delta$. The value of $U = 2.66 \times 10^{-10}$ eV is determined for the (1S) exciton level to be located at 10 meV less than the band-gap energy $E_g = 1.5$ eV with a cut-off wavenumber $k_c = 1.36 \times 10^9$ m$^{-1}$. The cavity mode for $q = 0$ is on resonance with the (1S) exciton level ($E_{\text{cav}} = E_g - 10$ meV), and the lower polariton level is formed at an energy of 10 meV below the cavity energy when using $\hbar g = 6.30 \times 10^{-7}$ eV. In addition, $\hbar\omega_{e,k} = \hbar\omega_{h,k} = \hbar^2k^2/2m + E_g/2$ ($m = 0.068\, m_0$, where $m_0$ is the free-electron mass) and $T = 10$ K are assumed with charge neutrality $\mu_e^B = \mu_h^B$. In this context, we note that the calculations are qualitative even though the parameters are determined as realistically as possible. For the other parameters, $\hbar\kappa = 100$ μeV and $\hbar\gamma = 4$ meV are used.

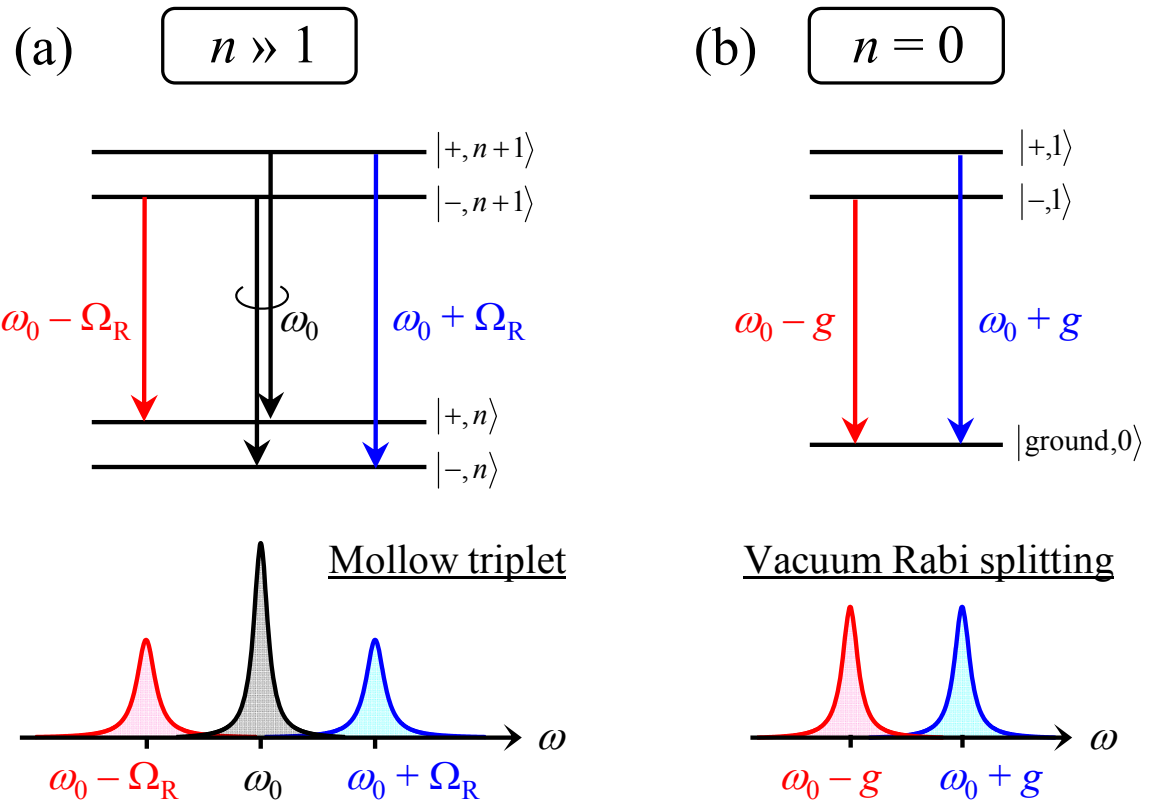

**Fig. S1.**
Dressed states and their spectra (**A**) for an atom with a strong field and (**B**) for an atom strongly coupled with a cavity.

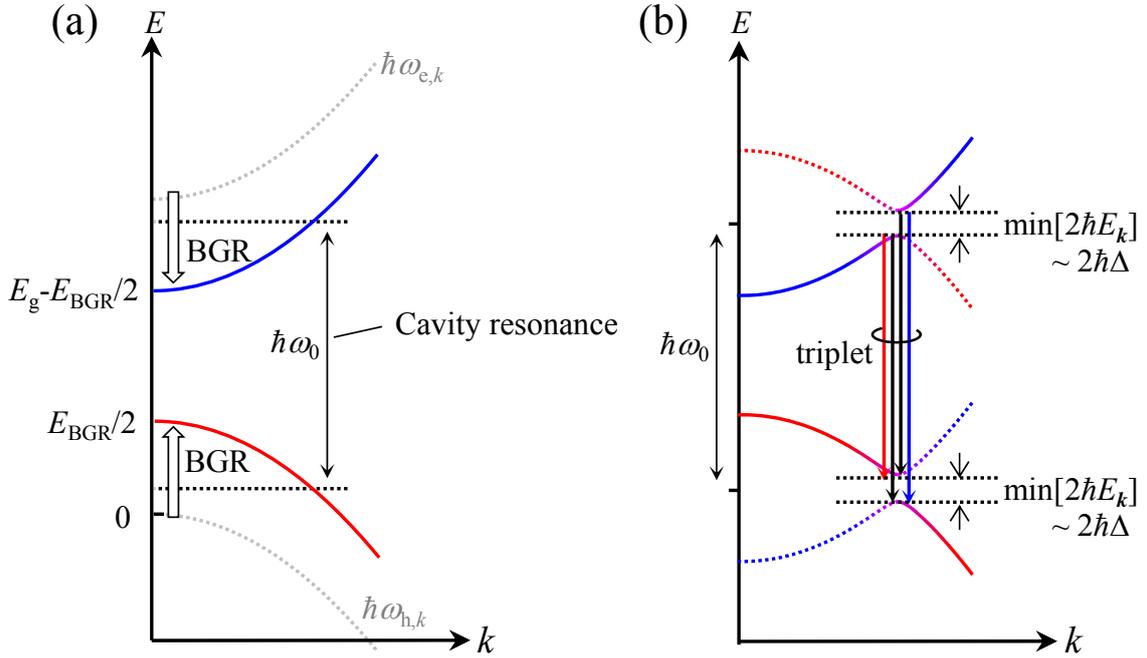

**Fig. S2.**
Energy diagram in a high-density regime. (**A**) The effects of the electron–electron and hole–hole Coulomb interactions are considered, but the dipole coupling to the cavity photons and the e–h attractive Coulomb interactions are neglected. (**B**) The effects of the dipole coupling and the e–h Coulomb interaction are also considered. In this case, the electron band (the solid blue curve) is mixed with the $+\hbar\omega_0$-shifted hole band (the dashed red curve). In the same manner, the hole band (the solid red curve) is mixed with the $-\hbar\omega_0$-shifted electron band (the dashed blue curve). Here, the contact potential $U'_{q-k} = U$ is assumed, and $\Delta_k = \Delta$.

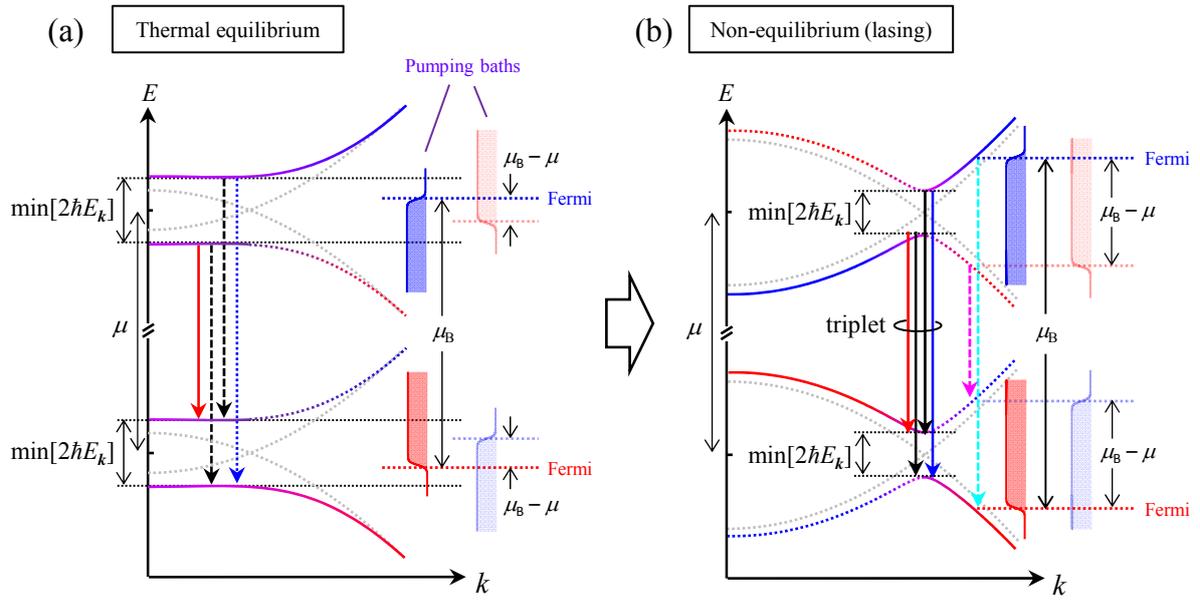

**Fig. S3.**
Energy diagram under thermal equilibrium and non-equilibrium conditions. These figures are the same as Fig. 3A and 3B in the main text.

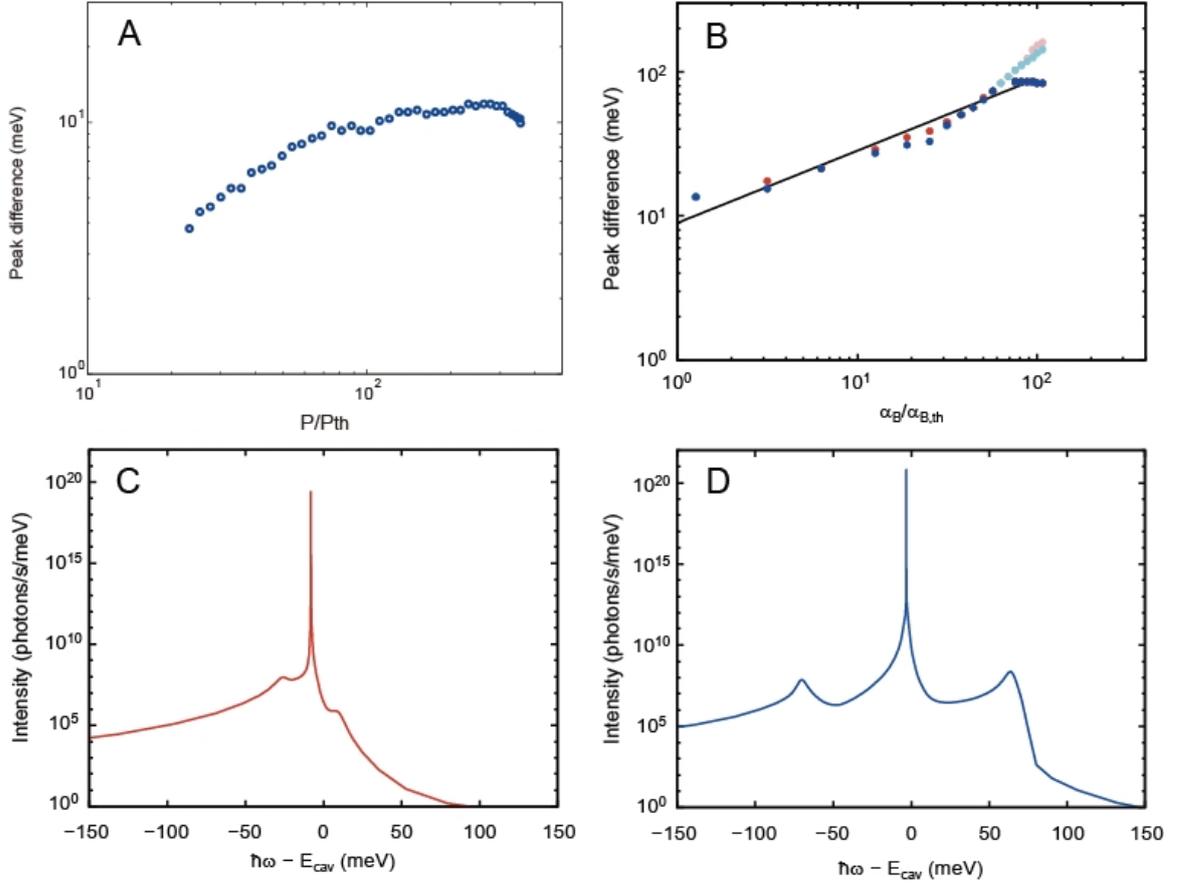

**Fig. S4.**
Experimental (panel (**A**)) and theoretical (panel (**B**)) results of the peak differences as a function of the pumping. Note that panel (A) is reprinted as Fig. 2 (D) while panel (B)-(D) are reprinted as Fig. 5.    In panel (B), the blue and red data points are the peak energies for the high- and low-energy peaks, respectively, which correspond to the transitions represented by the blue and red arrows in Fig. S3. In the lasing condition, the additional data points in aqua and pink correspond to the aqua and pink transitions in Fig. S3, respectively. The black lines of panels (A) and (B) are proportional to the square root of the pump power $P$ normalized to $P_{th}$ and $\alpha_B \equiv \mu_B - E_{LP}$ normalized to $\alpha_{B,th} \equiv \mu_{B,th} - E_{LP} = 1.64$ meV with the lower polariton energy $E_{LP}$ with zero detuning, respectively. The corresponding theoretical treatment is in Section S3. (**C** and **D**) Calculated photoluminescence spectrum $I_{SS}^{inc}(\omega, \boldsymbol{q} = 0)$ at $\mu_B - E_{LP} = 5$ meV ((C), low excitation density) and 80 meV ((D), high excitation density) indicated by arrows in Fig. 4B in the main text. The experimental peak differences are an order of magnitude smaller than the calculated values. This discrepancy may be caused by dephasing and polariton–polariton scattering processes not taken into account in our theory.

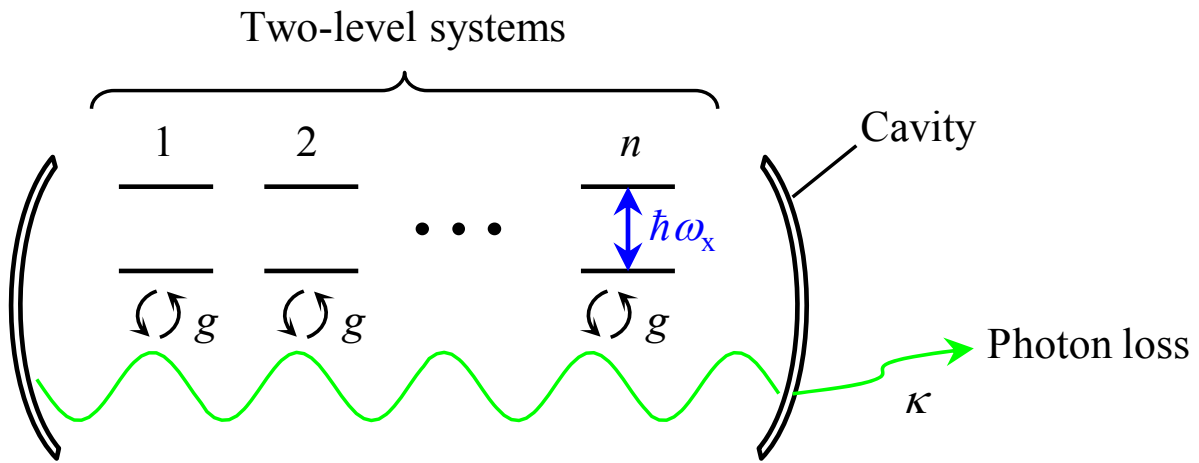

**Fig. S5.**
Schematic diagram of the system studied in this section (Section S3), showing $n$ equivalent two-level systems coupled with the cavity.

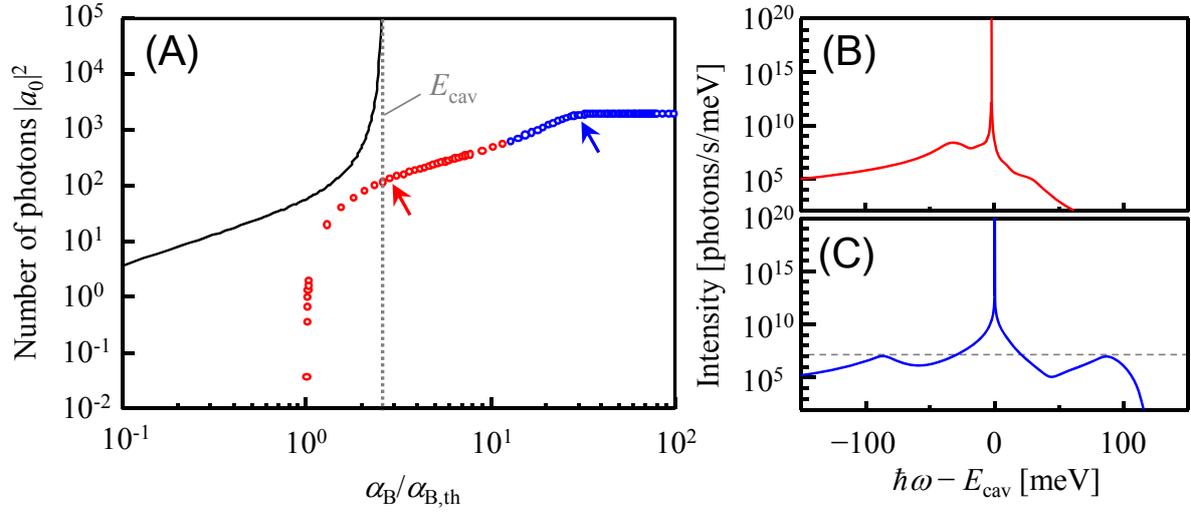

**Fig. S6.**
Results calculated with the TLS model. (**A**) The number of photons inside the cavity as a function of the pumping parameter $\alpha_B \equiv \mu_B - E_{LP}$ normalized to $\alpha_{B,th} \equiv \mu_{B,th} - E_{LP} = 3.88$ meV. Here, red data points and curves are used to indicate when the system can still be regarded as in a quasi-equilibrium state, and otherwise blue plot elements are used. For comparison, a result for the thermal equilibrium limit is also displayed by the solid black line ($\kappa = 0$ and $\gamma = 0^+$). The gray dotted line represents the bare cavity energy $E_{cav}$ (= $\hbar\omega_x$). (**B** and **C**): Calculated photoluminescence spectra $I_{SS}^{inc}(\omega, q = 0)$ at $\alpha_B = 20$ meV (B) and 120 meV (C), which correspond to the data points indicated by the arrows in panel (A). The gray dashed line in panel (C) is a guide for the eyes to help compare the sideband peak intensities.

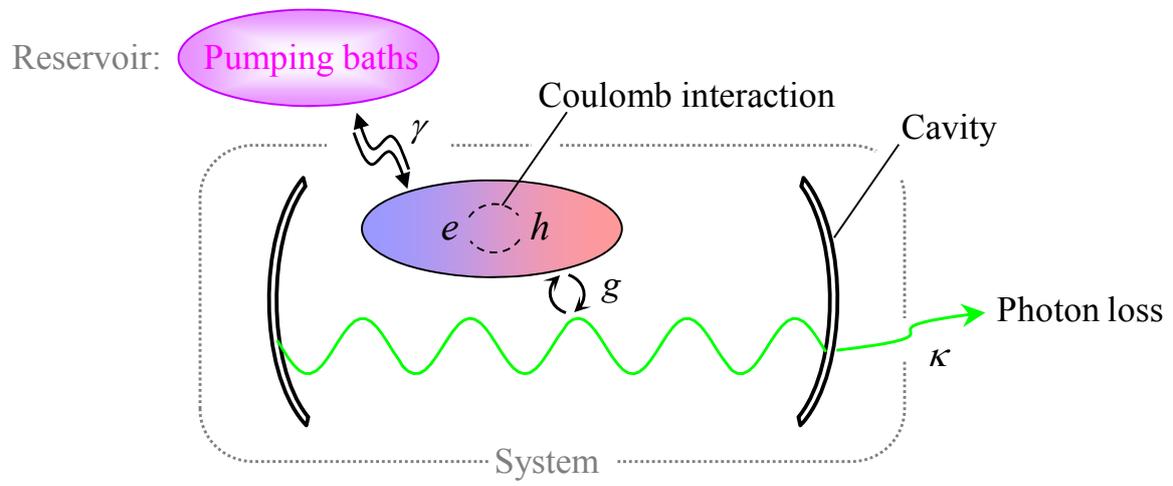

**Fig. S7.**
The model of the electron–hole–photon system.

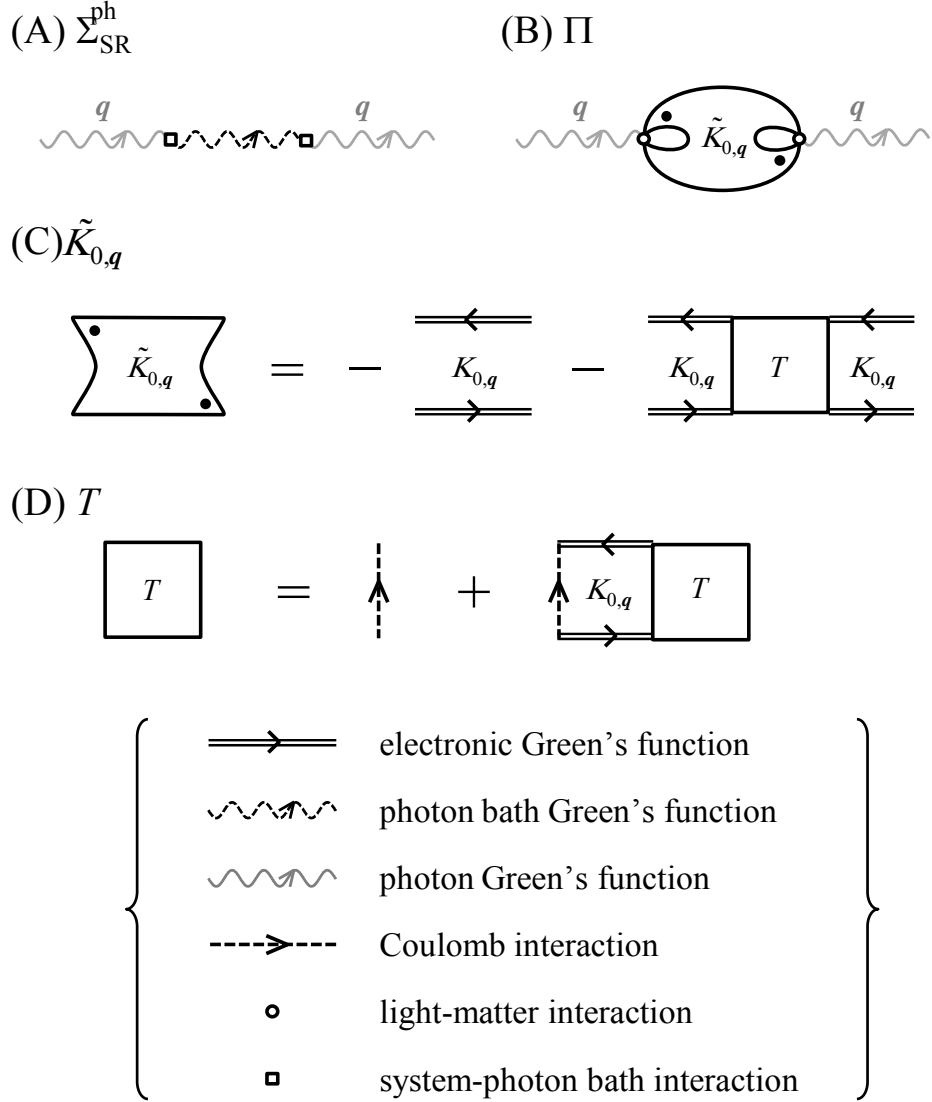

**Fig. S8.**
Self-energy diagrams with the *T*-matrix approximation. (**A**) $\Sigma_{SR}^{ph}$ and (**B**) $\Pi$ are the self-energies due to the system–reservoir coupling and the light–matter coupling, respectively. (**C**) $\tilde{K}_{0,q}$ and (**D**) $T$ are the two-particle Green's function and $T$ matrix, respectively. The double and wavy-dashed lines represent the electronic Green's function and photon-bath Green's functions, respectively. The gray wavy line represents the photon Green's function that will be connected to the self-energy diagrams. The dashed line corresponds to the Coulomb interaction.

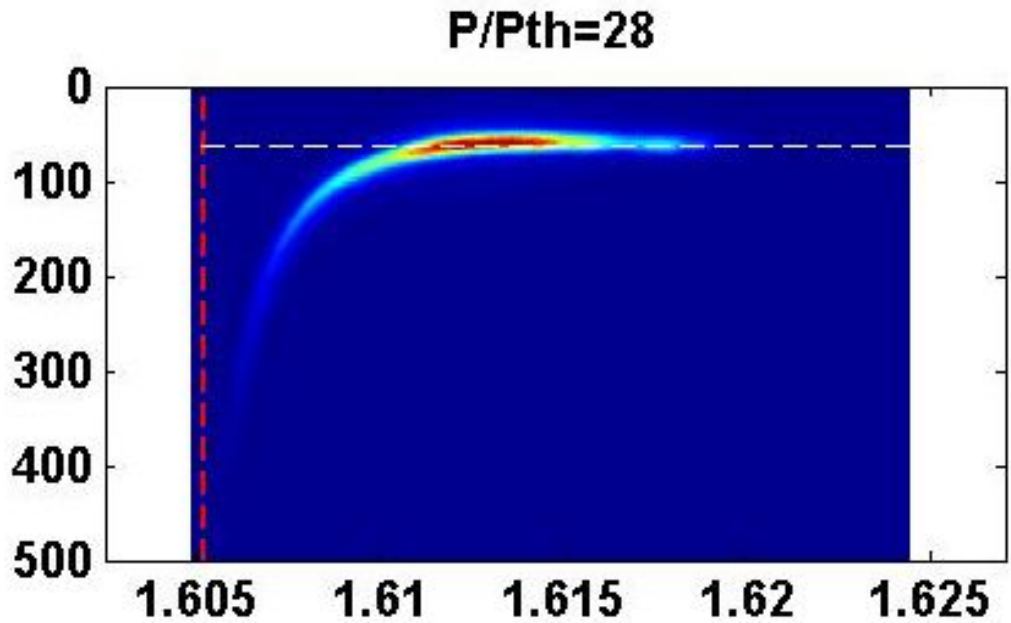

**Fig. S9.**
Energy relaxation in a high pump power regime ($28P_{th}$). Though the pump laser power is lower than that of Fig. 2A ($75P_{th}$), the behaviour of the relaxation into lower polariton energy far below $P_{th}$ shown by red dotted line is the same. At this pump power, the high-energy peak is still close to and not completely separated from the main peak as also shown in Fig. 2C.

**References**
43. Szymanska, M. H. Keeling, K. and Littlewood, P. B., Mean-field theory and fluctuation spectrum of a pumped decaying Bose-Fermi system across the quantum condensation transition. *Phys. Rev. B* **75**, 195331 (2007).
44. Keeling, J. Szymanska, M. H. and Littlewood, P. B., *Optical Generation and Control of Quantum Coherence in Semiconductor Nanostructures*, G. Slavcheva, P. Roussignol, Eds. (Springer, Berlin, 2010), *chap. 12*.
45. Wouters, M. and Carusotto, I. Excitations in a Nonequilibrium Bose-Einstein Condensate of Exciton Polaritons. *Phys. Rev. Lett.* **99**, 140402 (2007).
46. Chow, W. W. *et al.*, Nonequilibrium Model for Semiconductor Laser Modulation Response. IEEE *J. Quantum Electron*. **38**, 402 (2002).
47. Rammer, J. *Quantum Field Theory of Non-equilibrium States* (Cambridge University Press, New York, 2007).
48. Kamenev, A. *Field Theory of Non-Equilibrium Systems* (Cambridge University Press, New York, 2011).